\documentclass[10pt,prl,aps,twocolumn,superscriptaddress,preprintnumbers]{revtex4-1}

\usepackage[utf8]{inputenc}
\usepackage{epigraph}
\usepackage{amsmath}
\usepackage{wrapfig}
\usepackage{epsfig}
\usepackage{amssymb}
\usepackage{graphicx}
\usepackage{slashed}
\usepackage{xcolor}
\usepackage{comment}
\usepackage{natbib}
\usepackage{enumitem}
\usepackage{tikz-cd,tikz}
\usepackage{float}
\usepackage{array,adjustbox,booktabs}
\usepackage{subcaption}
\usepackage{amsfonts}
\usepackage{physics}
\usepackage{hyperref}

\hypersetup{
  linktocpage,
  colorlinks  = true, %Colours links instead of ugly boxes
  urlcolor    = blue, %Colour for external hyperlinks
  linkcolor   = blue, %Colour of internal links
  citecolor   = purple %Colour of citations
}

\usepackage[a4paper]{geometry}

\DeclareMathAlphabet\mathbfcal{OMS}{cmsy}{b}{n}

\usetikzlibrary{decorations.markings}
\tikzset{middlearrow/.style={
		decoration={markings,
			mark= at position 0.5 with {\arrow{#1}} ,
		},
		postaction={decorate}
	}
}

\def\be{\begin{equation}}
\def\ee{\end{equation}}

\def\cS{  {\cal S}  }

\def\cK{  {\cal K}  }

\def\cN{{\mathcal N}}
\def \cO{\mathcal{O}}
\def\l{\ell}

\DeclareMathOperator{\sgn}{sgn}

\def\ulmb{{\underline{\smash\lambda}}}

\def\umu{{\underline{\smash\mu}}}

\def\cO{{\mathcal{O}}}

\newcommand{\nocontentsline}[3]{}
\let\origcontentsline\addcontentsline
\newcommand\stoptoc{\let\addcontentsline\nocontentsline}
\newcommand\resumetoc{\let\addcontentsline\origcontentsline}

\usepackage{relsize}

\begin{document}

\title{
\makebox[\textwidth][c]{ \fontsize{11}{8}\selectfont
A formula for the block expansion in free CFTs 
and applications to ${\cal N}=4$ SYM at strong coupling
}}
\author{F. Aprile}
\affiliation{Departamento de F\'isica Te\'orica \& IPARCOS, Facultad de Ciencias F\'isicas, Universidad Complutense, 28040 Madrid}
\author{J. M. Drummond}
\affiliation{School of Physics and Astronomy, University of Southampton, Highfield,  SO17 1BJ}
\author{P. J. Heslop}
\affiliation{Department of Mathematical Sciences, Durham University, Durham, DH1 3LE}
\author{M. Santagata}
\affiliation{Department of Physics, National Taiwan University, Taipei 10617, Taiwan}

\begin{abstract}
\noindent 

An explicit analytic formula is presented that
computes the conformal (super-)block 
decomposition of any free scalar or half-BPS diagram
in 1d, 2d or 4d CFTs, with various supersymmetries, including none.
We prove our formula by exploiting a connection between conformal blocks 
and symmetric polynomials. Then we give a direct application of our result to the 
study of four-point correlators in ${\cal N}=4$ SYM at strong coupling. In particular, 
we give a CFT proof of the tree-level Witten diagram representation of 
$\langle {\cal O}_2^2{\cal O}_2^2 {\cal O}_q{\cal O}_q\rangle$ on AdS$_5\times$S$^5$, 
providing new and highly non-trivial checks of the AdS/CFT correspondence.
Our method works for a more general class of multi-particle correlators 
and can be used to bootstrap new results at strong coupling.

\end{abstract}

\maketitle

\stoptoc

%\section{Introduction}

In a conformal field theory (CFT) correlation functions have a 
convergent series expansion rooted in the properties of the Operator Product Expansion (OPE). 
The building blocks of such an expansion are 
the conformal blocks. However, often one faces the opposite task: reconstruct the OPE data from 
the conformal block expansion of a number of correlation functions. 
The simplest instance of this is when 
the CFT has a limit to free or generalized free fields, 
which happens in many cases at both weak and strong coupling. 
To specify the problem, take four scalar operators ${\cal O}_{i=1,\ldots 4}$, 
and consider all possible diagrams built out of free massless propagators. 
Say $G$ is one such diagram, its block decomposition is given by the equation, 
\begin{equation}\label{CFT_equa}
\begin{array}{c}
\begin{tikzpicture} [scale=1.1]
\def\shift{.35}

\def\latoxuno{-.35}
\def\latoxdue{-.38+0.8}
\def\latoyuno{.3}
\def\latoydue{-.25}

\draw (\latoxuno-.03, \latoyuno) --  (\latoxuno-.03, \latoydue);
\draw (\latoxuno+.03,\latoyuno) --  (\latoxuno+.03, \latoydue);
\draw (\latoxuno, \latoyuno) --  (\latoxdue, \latoyuno);
\draw (\latoxuno, \latoydue+.03) --  (\latoxdue, \latoydue+.03);
\draw (\latoxuno, \latoydue-.03) --  (\latoxdue, \latoydue-.03);
\draw (\latoxdue, \latoyuno) --  (\latoxdue, \latoydue);
\draw (\latoxdue, \latoyuno) --  (\latoxuno, \latoydue);

    \draw[fill=white!60,draw=white] (\latoxuno, \latoyuno) circle (2.5pt);
    \draw[fill=white!60,draw=white] (\latoxuno, \latoydue) circle (2.5pt);
    \draw[fill=white!60,draw=white] (\latoxdue, \latoyuno) circle (2.5pt);
    \draw[fill=white!60,draw=white] (\latoxdue, \latoydue) circle (2.5pt);

   \draw(\latoxuno-.1, \latoyuno+.1) node[scale=.8] {${\cal O}_1$};
   \draw(\latoxuno-.1, \latoydue-.1) node[scale=.8] {${\cal O}_2$};
   \draw(\latoxdue+.1, \latoyuno+.1) node[scale=.8] {${\cal O}_4$};
   \draw(\latoxdue+.1, \latoydue-.1) node[scale=.8] {${\cal O}_3$};

%-----------------------------------------------------------

\def\latoxuno{2.15+.55}
\def\latoxdue{3.75+.55}

  \draw(\latoxuno,\latoyuno) node[scale=.8] {${\cal O}_1$};
  \draw(\latoxuno, \latoydue) node[scale=.8] {${\cal O}_2$};
   \draw(\latoxdue+.05, \latoyuno ) node[scale=.8] {${\cal O}_4$};
  \draw(\latoxdue+.05, \latoydue) node[scale=.8] {${\cal O}_3$};

  \draw (\latoxuno+.2, \latoyuno) -- (\latoxuno+.5, 0.05 ) -- (\latoxuno+.2,\latoydue) ;
  
  \draw (\latoxuno+.8,.3) node[scale=.8] {$O_{\gamma,\ulmb}$};
  \draw (\latoxuno+.5, 0.05 ) -- (\latoxdue-.5, 0.05 ); 
  
   \draw (\latoxdue-.2,  \latoyuno ) -- (\latoxdue-.5, 0.05 )-- (\latoxdue-.2, \latoydue );

  \draw (\latoxdue+.4,0) node {$\displaystyle \Bigg],$};
  \draw (\latoxuno-.3,0) node {$\displaystyle \Bigg[$};

\draw (2.15-.7,-.1) node[scale=.9] {$\displaystyle=\sum_{\gamma,\ulmb} {M}_{G,\gamma,\ulmb}$};

\draw[thick, gray] (-.9,0.045) node[scale=.9] {$\displaystyle G=$};

   \end{tikzpicture} 
 \end{array}
\end{equation}
where the symbols on the RHS depict  
a sum over blocks, one for each exchanged primary operator 
$O_{\gamma,\ulmb}$ in both the $\cO_1\cO_2$ and $\cO_3\cO_4$ OPEs, and
the $M_{G,\gamma,\ulmb}$ are the coefficients that
we want to determine as a function of the quantum numbers, denoted
by $\gamma,\ulmb$. 
To achieve our goal of reconstructing OPE data, we want to determine
the $M_{G,\gamma,\ulmb}$
for all choices of  $G$,  in a number of different theories.
The theories we shall study include 4d scalar theories,
superconformal theories with $\cN=2,4$ supersymmetry,  
chiral 2d and 1d superconformal field theories, and more generally, 
all theories that have an $SU(m,m|2n)$ superconformal group. We 
deal with all of them in a unified manner, i.e.~for all $m,n$, 
by using the analytic superspace formalism~\cite{GIKOS,Howe:1995md,Heslop:2002hp,Doobary:2015gia},
which in particular gives a single expression for the (super-)conformal 
blocks. 
Now, a key outcome of the formalism is that the coefficients 
$M_{G,\gamma,\ulmb}$ can be shown to be theory-independent \cite{Doobary:2015gia}.
Using this independence, one can solve for the $M_{G,\gamma,\ulmb}$ of interest, in any of the theory mentioned above,
by embedding eq.~\eqref{CFT_equa} 
in a non-supersymmetric generalization of free theory where $n=0$ and $m$ is chosen appropriately. 
In this theory the blocks are simple generalizations of the 
standard conformal blocks written in \cite{Dolan:2000ut}. 
Then, by matching both sides of \eqref{CFT_equa}, 
and guesswork, one can gradually 
build up an \emph{explicit} formula
for the $M_{G,\gamma,\ulmb}$, which we will present as the main result of this letter.

On the other hand, the $(m,n)$ conformal blocks exhibit a rich mathematical structure \cite{Heckman_1,Heckman_2,Isachenkov:2016gim,Isachenkov:2017qgn,Schomerus:2021ins}
suggesting that a first principle proof of our formula for $M_{G,\gamma,\ulmb}$ ought to exist. 
The key in this case comes from \cite{Doobary:2015gia,Aprile:2017xsp,Aprile:2021pwd}
where it was shown that eq.~\eqref{CFT_equa} 
can be mapped to a Cauchy identity formula in the theory of BC symmetric polynomials.
Under this map, the block coefficients $M_{G,\gamma,\ulmb}$ are BC Jacobi 
polynomials evaluated at  $\{0,1\}$, 
where the number of 0's and 1's depends on the diagram $G$.
BC Jacobi polynomials, and hence the $M_{G,\gamma,\ulmb}$,  can be defined through 
a determinant formula \cite{Okounkov_Olshanski}, and from there,  
an \emph{elementary proof} of our formula follows, that 
does not rely on detailed knowledge of the blocks.

As a direct application of our result, we revisit the analytic bootstrap of tree-level correlators 
in ${\cal N}=4$ SYM at strong coupling, focusing on the  $SU(N)$ theory and the new class of correlators with two half-BPS
 single and two double-particle insertions. Our strategy is to compute data in the protected 
 sector, from the free theory, and use it as an input at strong coupling. In particular, we 
 will show how to use our formula to readily solve the mixing between protected operators and stringy 
states at the unitarity bound, which is crucial already at tree level or $O(1/N^2)$.   
We will carry out this analysis
for the first non-trivial family of such correlators,
$\langle {\cal O}_2^2{\cal O}_2^2 {\cal O}_q{\cal O}_q\rangle$ \cite{Aprile:2024lwy}. 
These have a tree-level $O(1/N^2)$ Witten diagram representation in AdS$_5\times$S$^5$ 
supergravity, whose validity will be successfully 
tested against our CFT results at strong coupling, providing a new and non-trivial test of AdS/CFT.

%==========================================================================
\section{Four-point function generalities} 
%==========================================================================

We consider external operators $\cO_p$ built from $p$ fundamental 
fields $\Phi_a(z)$
\begin{align}\label{O}
    \cO_p(z)=C^{[{\mathcal O}_p]}_{a_1 \dots a_p}:\Phi_{a_1}(z)\dots\Phi_{a_p}(z):\ .
\end{align}
The choice of $C^{[{\mathcal O}_p]}$ defines the operator, and depends on the theory. For example, in
the ${\cal N}=0$ theory of free scalars, we can have ${\cal O}_2=
\vec{\Phi}.\vec{\Phi}$, while in ${\cal N}=4$ SYM with $SU(N)$ gauge group we can have ${\cal O}_2=\tr\Phi^2$, 
and more generally the $C_{a_1 \dots a_p}$ can be taken to be any gauge invariant combination of generators. 
The field $\Phi_a(z)$ is defined through the analytic superspace formalism,  
and here $z$ are coordinates on $Gr(m|n,2m|2n)$~\cite{Howe:1995md,Heslop:2002hp,Doobary:2015gia}.
In free theory, $\Phi(z)$ has the two point function
\begin{align}
    \langle \Phi_a(z_i) \Phi_b(z_j) \rangle=\delta_{ab}\,g_{ij}:=\delta_{ab}\,\text{sdet} (z_j{-}z_i)^{-1}\,,
\end{align}
and the 4pt functions of interest are obtained by Wick's theorem:
\begin{align}\label{intro_4pt}
\langle \mathcal{O}_{p_1} \mathcal{O}_{p_2} \mathcal{O}_{p_3} \mathcal{O}_{p_4}\rangle
              = \ \sum_{G} a_{G} \times G \quad;\quad G=\prod_{i<j} g_{ij}^{b_{ij}}\,.
               \end{align}
The $a_{G}$ are determined by Wick contractions in terms of the $C_{a_1...a_p}$ 
coefficients. The diagram $G$ is specified by
the positive integers $b_{ij}$  such that $\sum_i b_{ij}=p_j$ with
$b_{ij}=b_{ji}$, $b_{ii}=0$. Pictorially,
\begin{equation}\label{intro_4pt1}
\!\!\!\!\!\!\!\!\begin{tikzpicture}[baseline,scale=0.4]
		\def\lato 	{4}
		\def\xuno	{2}
		\def\yuno	{2}
		\def\xdue {2}
		\def\ydue {2-\lato}
		\def\xtre {2+\lato}
		\def\ytre {2-\lato}
		\def\xquat {2+\lato}
		\def\yquat {2}

\draw (0.3,-.5) 			 node[left] {$\displaystyle G\ =\ \prod_{i<j} g_{ij}^{b_{ij}}=$};     
%
%bundles		
		\foreach \x in {-1,0,1,2}
		\draw[thick]    (\xuno,\yuno+\x*.1 ) --  (\xquat,\yquat+\x*.1 );
		\foreach \x in {-1,0,1}
		\draw[thick]    (\xuno+\x*.1,\yuno) --  (\xdue+\x*.1,\ydue );				
		\foreach \x in {-1,1}
		\draw[thick]    (\xdue,\ydue+\x*.07) --  (\xtre,\ytre+\x*.07 );	
		\foreach \x in {-1,1}
		\draw[thick]    (\xtre+\x*.07,\ytre) --  (\xquat+\x*.07,\yquat );	
		\foreach \x in {-1,0}
		\draw[thick]    (\xdue-\x*.1,\ydue+\x*.1) --  (\xquat-\x*.1,\yquat +\x*.1);		
		\foreach \x in {0}
		\draw[thick]    (\xtre-\x*.1,\ytre+\x*.1) --  (\xuno-\x*.1,\yuno +\x*.1);			
		\draw (\xuno-.25,\yuno+.5) 			 node[left] {$\mathcal{O}_{p_1}$};
		\draw (\xdue-.25,\ydue-.5)  			node[left] {$\mathcal{O}_{p_2}$};
		\draw (\xtre+.25,\ytre-.5)  			node[right] {$\mathcal{O}_{p_3}$};
		\draw (\xquat+.25,\yquat+.5)  		node[right] {$\mathcal{O}_{p_4}$};
		\draw (\xuno+.1,\yquat-1.2) 			node[left] 		{$\scriptstyle b_{12}$};
		\draw (\xtre-.1,\yuno-1.3) 			node[right] 	{$\scriptstyle b_{34}$};
		\draw (\xuno+.5,\yquat+.7) 			node[right] 	{$\scriptstyle b_{14}$};
		\draw (\xuno+.6,\ytre-.5) 			node[right] 	{$\scriptstyle b_{23}$};	
		\draw (\xuno+.2,\ytre+1.8) 			node[right] 	{$\scriptstyle b_{24}$};		
		\draw (\xuno+2.6,\ytre+1.4) 			node[right] 	{$\scriptstyle b_{13}$};		
%			
%points		
%		
		\draw[fill=red!10,draw=black] (\xuno,\yuno) circle (.4cm);
		\draw[fill=red!10,draw=black] (\xdue,\ydue) circle (.4cm);
		\draw[fill=blue!10,draw=black] (\xtre,\ytre) circle (.4cm);
		\draw[fill=blue!10,draw=black] (\xquat,\yquat) circle (.4cm);
	   \end{tikzpicture}
\end{equation}
Then, we can always arrange $p_{43},p_{12} \geq 0$ (where $p_{ij}\equiv p_i{-}p_j$) by 
swapping operator positions, whilst maintaining the $(12)(34)$ OPE channel \eqref{CFT_equa}.
In a generalized free theory, we will have
the same structure as in \eqref{intro_4pt1}. \vspace{2mm}

Let $\mathcal{P}=g_{12}^{ \frac{1}{2}(p_1+p_2) } g_{34}^{\frac{1}{2}(p_3+p_4) } 
\left({ g_{14}  }/{ g_{24} }\right)^{\frac{1}{2} p_{12}} 
\left( { g_{14} }/{ g_{13} } \right)^{ \frac{1}{2}p_{43} }\,,$
and rewrite each diagram $G$ as
\begin{equation}\label{defgammak}
G_{p_i;\gamma,k}=\prod_{i<j} g_{ij}^{b_{ij}}  = 
{\cal P}[g_{ij}]\times 
\left(  \frac{ g_{13} g_{24} }{ g_{12} g_{34} }\right)^{\!\frac{\gamma}{2} }\,  
\left(\frac{ g_{14}  g_{23} }{ g_{13} g_{24} }\right)^{\!\! k}
\end{equation}
with $\gamma$ and $k$ constrained by $b_{ij}\geq0$ to take the values 
\begin{align}  
\gamma=\gamma_{\text{min}},\gamma_{\text{min}}+2,\ldots,\gamma_{\text{max}};
\quad 
k=0,1,\ldots,\tfrac12(\gamma-\gamma_{\text{min}})\notag\\[.2cm]
\gamma_{\text{min}}= \max(p_{12},p_{43})\quad;\quad 
\gamma_{\text{max}}=\min( p_1{+}p_2, p_3{+}p_4 )\,.\notag
\end{align} 
Note that $\gamma=b_{13}{+}b_{14}{+}b_{23}{+}b_{24}$ 
counts the number of propagators on the vertical cut of  $G$. 
E.g.~$\gamma=4$ in \eqref{CFT_equa}.\vspace{1mm}

Consider now the OPEs: $\cO_{p_1} \cO_{p_2}\sim \sum \cO $ and $\cO_{p_3}\cO_{p_4} \sim \sum \cO'$. 
In the free theory  the exchanged primaries are explicit \cite{Doobary:2015gia}. They are characterised  
by $\gamma$ and a Young diagram $\ulmb$, and given by 
\begin{equation}\label{OPEfree}
\begin{array}{l}
\cO_{\gamma,\ulmb}\, \simeq  \Phi^{p_1-b_{12}} (\partial^{|\ulmb|}) \Phi^{p_2-b_{12}} +... \,,
\quad \gamma=p_1{+}p_2{-}2b_{12}\\[.2cm] 
\cO'_{\gamma,\ulmb}\simeq \Phi^{p_3-b_{34}} (\partial^{|\ulmb|})\Phi^{p_4-b_{34}}+...\,,\quad \gamma=p_3{+}p_4{-}2b_{34}\,,
\end{array}
\end{equation}
where the ellipses denote terms with derivatives redistributed (making the result primary) and $\ulmb$ 
gives the symmetrisation of the indices of the derivatives \cite{footnote1}. The dictionary between $\gamma,\ulmb$ and 
the more standard CFT notation is known and reviewed in the supplementary material (see e.g. (\ref{tableQN1})).
Note that  $\gamma$ plays a dual role: it counts 
the number of $\Phi$s constituting the exchanged operators 
in the OPE in~\eqref{OPEfree}
and it specifies the free 
theory diagram in~\eqref{defgammak}.

The decomposition in blocks of a 4pt correlator is
\begin{align}
\label{intro_formula_blockdeco}
\langle \mathcal{O}_{p_1} \mathcal{O}_{p_2} \mathcal{O}_{p_3} \mathcal{O}_{p_4} \rangle  = 
 \mathcal{P}[g_{ij}] 
 %\sum_{\gamma, \ulmb} 
 \sum_{\gamma,\ulmb}
{A}^{(m,n)}_{\gamma,\ulmb }
B_{\gamma,\ulmb}^{(m,n)}(z_i)
\end{align}
where $B^{(m,n)}_{\gamma,\ulmb}$ are the conformal blocks, and the $A_{\gamma,\ulmb}$ contain information about the OPE,
\begin{equation} \label{blockcoeff}
{A}_{\gamma,\ulmb } = \sum_{i,j}
C_{\cO_{p_1}\cO_{p_2} \cO^{(i)}_{\gamma,\ulmb}}\  
%C_{p_1p_2O^{(i)}_{\gamma,\ulmb} }
{G}^{ij}_{\gamma,\ulmb}\,\, C_{\cO_{p_3}\cO_{p_4}\cO^{(j)}_{\gamma,\ulmb}}. 
%C_{p_3p_4O^{(j)}_{\gamma,\ulmb}} 
\end{equation}
The $\cO^{(i)}_{\gamma,\ulmb}$ are a basis for the operators in \eqref{OPEfree}, ${G}^{ij}$ is the 
inverse two-point function metric in this basis, and the $C$s are three-point couplings. 
$B_{\gamma,\ulmb}$ and ${A}_{\gamma,\ulmb}$ depend on $p_{12}$, $p_{43}$, 
but sometimes we will omit them to simplify notation.
%
%

%==========================================================================
\paragraph{\bf Theory independence of the $M$aster coefficients.}
%==========================================================================
%
Performing the block expansion of \eqref{intro_4pt} it is useful to split it 
into separate pieces, one for each diagram $G$, and by doing so define
\begin{align}\label{graphbd}
G_{p_i;\gamma,k}= {\cal P}[g_{ij}] \sum_{\ulmb} M_{k,\gamma,\ulmb}B_{\gamma,\ulmb}^{(m,n)}\,.
\end{align}
This is formula \eqref{CFT_equa} given in a more precise notation.
Clearly, from \eqref{intro_4pt} and \eqref{intro_formula_blockdeco}, the $A_{\gamma,\ulmb}$ in \eqref{blockcoeff} are 
\begin{align}\label{A=aM}
%$
A_{\gamma,\ulmb} = \sum_k a_{G_{\gamma,k}} M_{k,\gamma,\ulmb}\,.
%$.\vspace{2mm}
\end{align}
Writing  $B_{\gamma,\ulmb}^{(m,n)}=
\left( { g_{13} g_{24} }/{ g_{12} g_{34} }\right)^{\!\frac{\gamma}{2}}\!\!\times F_{\gamma,\ulmb}^{(m,n)}$ 
as in~\cite{Doobary:2015gia},  eq.~\eqref{graphbd} becomes
\begin{equation}\label{def_calA_k}
   \left(\frac{ g_{14}  g_{23} }{ g_{13} g_{24} }\right)^{\!\! k} =
 \sum_{ \ulmb} M_{k,\gamma,\ulmb} F_{\gamma,\ulmb}^{(m,n)}\,.
\end{equation}
It will be useful to view both sides of \eqref{def_calA_k}
as class functions of the $(m+n)^2$ cross-ratio supermatrix
$z=z_{12}z_{24}^{-1}z_{43}z_{31}^{-1}$ 
$=\text{diag}(x_{i=1..m},y_{j=1..n})$.
In particular, 
\begin{align}\label{sd1-z}
    \frac{ g_{14}  g_{23} }{ g_{13} g_{24} } = \text{sdet} (1{-}z) = \frac{\prod_{j=1}^n(1{-}y_j)}{\prod_{i=1}^m(1{-}x_i)}\,.
\end{align}
In supersymmetric theories $m,n\neq0$, and 
the $F^{(m,n)}$
depend on the type of multiplet through $\gamma$ 
and the data of a hook shaped Young diagram $\ulmb$,  
as depicted below\vspace{-2mm}
\begin{equation}
\begin{array}{c}
\begin{tikzpicture}[scale=.9]

\draw [fill=gray!20,draw=white](0,0) rectangle (1.8,-.2);
\draw [fill=gray!20,draw=white](0,-.2) rectangle (1.3,-.4);

\draw [fill=gray!20,draw=white](0,0) rectangle (.2,-1.8);
\draw [fill=gray!20,draw=white](.2,0) rectangle (.4,-1.2);

\draw[draw=white] (0,-.2) -- (.4,-.2);
\draw[draw=white] (0,-.4) -- (.4,-.4);

\foreach \x / \y   in  { 2/0,  2/-.6 }  %.6/-2
		\draw[dashed] (0, \y) -- (\x,\y);

\foreach \x / \y   in  { 0/-2,  .6/-2} % 2/-.6 
		\draw[dashed] (\x, 0) -- (\x,\y);
        
\draw[ latex reversed-latex reversed,  gray] (-.275,0) -- (-.275,-.6);
\draw[ latex reversed-latex reversed,  blue!50] (-.125,0) -- (-.125,-1.8);
\draw[ latex reversed-latex reversed, gray] (0,.2) -- (.6,.2);

\draw [dotted, gray]   (0,-1.8)--(3.2,-1.8);
\node[scale=.9] at (3.2,-1.65) {$\lambda_{\beta}$};

\node[scale=.9] at (-.5,-.3) {$m$};
\node[scale=.9] at (+.3,+.4) {$n$};

\node[scale=.9] at ( +3.7,-.65) {$\lambda_{m+1}\leq n$};
\draw [dotted, gray]   (0,-.8)--(3.2,-.8);

\end{tikzpicture}
\end{array}\vspace{-1mm}
\end{equation}
A Young diagram of this type has the first $m$ rows and the first $n$ columns unbounded, 
therefore the total number of rows of $\ulmb$ is not fixed, but varies depending 
on the R-symmetry rep of the exchanged primary. 
For fixed charges $p_{i}$ the maximum height of $\ulmb$ is given by $\beta=\frac{1}{2}(\gamma-\gamma_{\rm min})$. 
The $F^{(m,n)}$ are known \cite{Doobary:2015gia}, however, as the $p_i$ increase, 
the supersymmetric block expansion becomes unfeasible. Now, a key feature of our $(m,n)$ 
formalism is that the $M_{k,\gamma,\ulmb}$ can be proven to be independent 
of the dimension and number of SUSY's \cite{Doobary:2015gia}. 
Then we can embed the $M_{k,\gamma,\ulmb}$ of interest in 
a non supersymmetric generalization of free theory, the $(\beta,0)$ theory,
and perform the block expansion in this theory. In this theory there are $\beta$ 
variables $x_i$, however the blocks are scalable, and simply given by the 
following compact  expression
\begin{align}\label{blockbosonic}
& F^{(\beta,0)}_{\gamma,\ulmb} =  \\
& \frac{\substack{\displaystyle \det \\[0.5mm] 1\leq i,r\leq\beta} \left( x_i^{\lambda_{r}{{+}}\beta{-}r} {}_2F_1\Big[\,^{\lambda_{r}+1-r+\frac{\gamma-p_{12}}{2}\,;\,
\lambda_{r}+1-r+\frac{\gamma-p_{43}}{2}}_{\rule{1.4cm}{0pt}2\lambda_{r}+2-2{r}+\gamma};x_i\Big] \right)}{\prod_{i<{r}}(x_i-x_{r})}\,
\notag
\end{align}
Implementing our strategy, and studying many cases, 
we were able to obtain a completely general formula for $M_{k,\gamma,\ulmb}$, correctly yielding the block coefficients for any free 
theory diagram as an analytic function of $\gamma, \ulmb, k, p_i$.

%==============================================================
\section{\bf Master formula}
%===============================================================

The coefficients $M^{\vec{p}}_{k,\gamma,\ulmb}$ are consistent with  the formula 
\begin{widetext}
\begin{equation}\label{Acoeffs}
M^{\vec{p}}_{k,\gamma,\ulmb}= \prod_{i=0}^{k-1}\!\frac{1}{i! (p_{+}{+}i)!} 
\!\!\prod_{i=0}^{\beta-k-1}\!\!\frac{1}{i! (p_{-}{+}i)!} \,\prod_{i=1}^{\beta}B_i
\sum_{S_k \subseteq S_\beta} \bigg[
\prod_{\substack{i < j \\ i,j \in S_k}}C_{ij}  
\prod_{\substack{\bar \imath< \bar \jmath \\ \bar \imath,\bar \jmath \in \overline{S}_k}} C_{\bar \imath \bar \jmath}
\prod_{i \in S_k} (\hat{\lambda}_i {+} p_{+})!
\prod_{\bar \imath \in \overline{S}_k} %\left(
(-1)^{\lambda_{\bar \imath}}(\hat{\lambda}_{\bar \imath} {+} p_{-})!%\right)\
\,\,\bigg]
\end{equation}
\end{widetext}
where  
$\ulmb=[\lambda_1,\ldots,\lambda_{\beta}]$ --  note the row lengths can be vanishing --   
then $\hat{\lambda}_i=\lambda_i{+}\beta{-}i$, and
\begin{equation}\label{BCcoeffdef}  
\!\!C_{ij}=(\hat{\lambda}_i {-}\hat{\lambda}_j)(\hat{\lambda}_i {+} \hat{\lambda}_j{+}1+ \hat{p})\ \ ;\ \  
B_i = \frac{(\hat{\lambda}_i{+} \hat{p})!}{(2\hat{\lambda}_i{+} \hat{p})!}\,,
\end{equation}
with $\hat{p}\equiv p_++p_-$, $p_{\pm}=\frac{1}{2}| p_{43}\pm p_{12}|$.
The sum in \eqref{Acoeffs} runs over subsets $S_k$ of 
$S_\beta= \{1,\ldots,\beta \}$ of size $k$. The complement 
of $S_k$ in $S_\beta$ is denoted with $\overline{S}_k$.

%=======================================================================
\paragraph{\bf Proof of the master formula.}
%=======================================================================
Although originally found by examination, formula \eqref{Acoeffs} can be proven.
The starting point is the ``superconformal Cauchy identity"  with $\theta=1$ presented in  
\cite{Aprile:2021pwd} which reads 
\begin{equation}\label{superCauchy2}
\frac{\,\prod_{\l,j}^{d,n} 
(1-Y_{\l} y_j)}{\!\!\prod_{\l,i}^{d,m} 
		(1-Y_{\l} x_i)} =  	
	\sum_{\ulmb } 
	 J^{(d)}_\ulmb(Y) \,
		{F}^{(m,n)}_{\gamma(d),\,\ulmb}(x,y)\,.
\end{equation}
Here $J^{(d)}$ are $d$-variable Jacobi polynomials \cite{Okounkov_Olshanski,Koornwinder}, 
and $\gamma(d)=2d + \hat{p}$. It is worth emphasizing that
\eqref{superCauchy2} does not rely on detailed knowledge of the blocks and 
can be derived only from general considerations about the supersymmetric uplift of known 
Cauchy identities for symmetric polynomials, as shown in \cite{Aprile:2021pwd}.
Now, comparing (\ref{def_calA_k},\ref{sd1-z}) with \eqref{superCauchy2} 
we see that if we set  $d=\beta$, and assign values $Y_{i=1,\ldots,k}=1$, we 
match both the CFT value of $\gamma$ in \eqref{def_calA_k} and the cross ratio in \eqref{sd1-z}.
Therefore the $M$ coefficients must also match, which means
\begin{equation}\label{AtoJ}
M_{k,\gamma,\ulmb}=J^{(\beta)}_{\ulmb}(Y;p_-,p_+)\Big|_{Y=[1^k,0^{\beta-k}]}\,.
\end{equation}
The next step is to derive an explicit formula for \eqref{AtoJ}. 
We shall use two ingredients. The first one is a 
determinantal expression for the Jacobi polynomials 
~\cite[(7.2)]{Okounkov_Olshanski}
given in terms of ordinary one-variable Jacobi polynomials
\begin{equation}\label{Jdet}
J^{(d)}_{\ulmb}=\frac{ \substack{\displaystyle \det \\[0.5mm] 1\leq i,\l\leq d} \bigg( J_{[\lambda_{i}+d-i]}(Y_\l) \bigg)
}{\prod_{i<\l} (Y_i-Y_\l) }\ .
\end{equation}
(For convenience we recall the definition of $J$ in the supplemental material.)
The second ingredient are the so called evaluation formulae for 
Jacobi polynomials in which $Y$ is either completely zero or unity, 
see e.g.~\cite{Koornwinder,opdam1989some}: 
\begin{align}
    \label{evaluationknown1}
    J_{[\mu_1..\mu_d]}^{(d)}({0};p_{-},p_{+})&= (-1)^{|\umu|} f({\mu_i{+}d{-}i}\,;p_-,p_+) \\
    \label{evaluationknown2}
    J_{[\mu_1..\mu_d]}^{(d)}({1};p_{-},p_{+})&= \ \ \ \ \  f({\mu_i{+}d{-}i}\,;p_+,p_-)
\end{align}
where we introduced
\begin{equation}\label{functionf}
    f(\mu_i\,;p_-,p_+)= \prod_{i=1}^{d}\frac{B_i \,( \mu_{i} {+} p_{-})! }{(i{-}1)! (p_{-}{+}i{-}1)!} \prod_{\substack{i< j \leq M  } } C_{ij}\,.
\end{equation}
Here the $B_i,C_{ij}$ are defined as in \eqref{BCcoeffdef}. 
At this point, the determinant expression \eqref{Jdet} and the evaluation formulae 
(\ref{evaluationknown1}-\ref{evaluationknown2}) imply the following identities
\begin{align}
\left.    \frac{\det \big(J_{\mu_i}(Y_{\l})\big)
}{\prod_{i<{\l}} (Y_i{-}Y_{\l})}\right|_{Y_{\l}=0} &=(-1)^{\frac12d(d-1)+\sum_i\mu_i} f(\mu_i;p_-,p_+) \,, \notag \\
\label{dJ1}
\left.    \frac{\det \big( J_{\mu_i}(Y_{\l})\big)
}{\prod_{i<{\l}} (Y_i{-}Y_{\l})}\right|_{Y_{\l}=1}& =f(\mu_i;p_+,p_-)\ .
\end{align}
A key point of these identities is that they are both 
true as function of $\underline{\mu}=[\mu_1,\ldots \mu_d]$ 
for any $d \times d$ dimension matrix on the LHS.
Finally, to evaluate \eqref{AtoJ} at $Y=[1^k,0^{\beta-k}]$ we start from the determinant 
expression in \eqref{Jdet} and split it into the product of $k \times k$ minors evaluated 
at $1$, times $(\beta-k)\times(\beta-k)$ minors evaluated at $0$ using the Laplace 
cofactor expansion. The denominator splits into three factors, but one of them 
equals unity and we obtain
\begin{widetext}
\begin{equation}\label{dJ2}
M_{k,\gamma,\ulmb} =  \sum_{S_k\subseteq S_\beta} \sgn(\{S_k,\bar S_k\})\  \times\  
\frac{{\det}_k\left(\Big[J_{[\lambda_{i}+\beta-{i}]}(Y_{\l})\Big]{\substack{\scriptscriptstyle i\in S_k \\[1mm]\scriptscriptstyle \l\leq k}}\right)}{\rule{0pt}{.3cm}\prod_{i<\l\leq k} (Y_i-Y_\l)}\Bigg|_{Y\rightarrow {1}} \times\ \ 
%\!\!\!\!\!\!\!\!
\frac{{\det}_{\beta-k}\left(\Big[J_{[\lambda_{i}+\beta-i]}(Y_{\l})\Big]{\substack{\scriptscriptstyle i\in \bar S_k\\[1mm]\scriptscriptstyle \l>k}}\right)}{\rule{0pt}{.3cm} \prod_{k<i<\l} (Y_i-Y_{\l})}\Bigg|_{Y\rightarrow {0}} \,.  
\end{equation}
\end{widetext}
Inserting~\eqref{dJ1} directly yields  the master formula~\eqref{Acoeffs} 
(after some manipulation of overall signs).

%=========================================================
\paragraph{\bf Basic examples.} 
%=========================================================
Focussing on external fields of charge ${p=2}$, we find diagrams with $\gamma=0,2,4$. 
The $\gamma=0$ diagram corresponds to the exchange of the identity, and it is trivial. 
For the $\gamma=2$ diagrams we find
\begin{align}\label{A2222_ini}
&
\!\!\!M_{1,2,[\lambda]}=  \frac{(\lambda!)^2}{(2\lambda)!}\,\quad,\quad\, M_{0,2,[\lambda]}= M_{1,2,[\lambda]} (-1)^{\lambda}\,.
\end{align}
The $\gamma=4$ diagrams are
\begin{equation}\label{gamma4diagrams}
\!\!\!\!\begin{array}{c}
\!\!\!\begin{tikzpicture}[scale=1.2]  
\def\shift{.35}

\def\latoxuno{-.35}
\def\latoxdue{-.38+0.8}
\def\latoyuno{.3}
\def\latoydue{-.25}

\draw (\latoxuno+.03, \latoyuno+.03) --  (\latoxdue-.03, \latoydue+.03);
\draw (\latoxuno+.03, \latoyuno-.03) --  (\latoxdue-.03, \latoydue-.03);
\draw (\latoxuno+.03, \latoydue+.03) --  (\latoxdue-.03, \latoyuno+.03);
\draw (\latoxuno+.03, \latoydue-.03) --  (\latoxdue-.03, \latoyuno-.03);

\draw[fill=white!60,draw=white] (\latoxuno-.075, \latoyuno-.075) rectangle (\latoxuno+.075, \latoyuno+.075);
\draw[fill=white!60,draw=white] (\latoxuno-.075, \latoydue-.075) rectangle (\latoxuno+.075, \latoydue+.075);
\draw[fill=white!60,draw=white] (\latoxdue-.075, \latoyuno-.075) rectangle (\latoxdue+.075, \latoyuno+.075);
\draw[fill=white!60,draw=white] (\latoxdue-.075, \latoydue-.075) rectangle (\latoxdue+.075, \latoydue+.075);

   \draw(\latoxuno-.1, \latoyuno+.1) node[scale=.8] {${\cal O}_2$};
   \draw(\latoxuno-.1, \latoydue-.1) node[scale=.8] {${\cal O}_2$};
   \draw(\latoxdue+.1, \latoyuno+.1) node[scale=.8] {${\cal O}_2$};
   \draw(\latoxdue+.1, \latoydue-.1) node[scale=.8] {${\cal O}_2$};

   \draw (\latoxdue+2.3,-0.1) node[scale=1] {${\displaystyle= g_{12}^2 g_{34}^2 \sum_{\ulmb} M_{0,4,[\lambda_1,\lambda_2]} B_{4,\ulmb}^{(m,n)}}$};

\end{tikzpicture}
\\[.2cm]
\!\!\!\begin{tikzpicture}[scale=1.2]  
\def\shift{.35}

\def\latoxuno{-.35}
\def\latoxdue{-.38+0.8}
\def\latoyuno{.3}
\def\latoydue{-.25}

\draw (\latoxuno, \latoyuno) --  (\latoxdue, \latoydue);
\draw (\latoxuno, \latoydue) --  (\latoxdue, \latoyuno);
\draw (\latoxuno, \latoyuno) --  (\latoxdue, \latoyuno);
\draw (\latoxuno, \latoydue) --  (\latoxdue, \latoydue);

\draw[fill=white!60,draw=white] (\latoxuno-.075, \latoyuno-.075) rectangle (\latoxuno+.075, \latoyuno+.075);
\draw[fill=white!60,draw=white] (\latoxuno-.075, \latoydue-.075) rectangle (\latoxuno+.075, \latoydue+.075);
\draw[fill=white!60,draw=white] (\latoxdue-.075, \latoyuno-.075) rectangle (\latoxdue+.075, \latoyuno+.075);
\draw[fill=white!60,draw=white] (\latoxdue-.075, \latoydue-.075) rectangle (\latoxdue+.075, \latoydue+.075);

   \draw(\latoxuno-.1, \latoyuno+.1) node[scale=.8] {${\cal O}_2$};
   \draw(\latoxuno-.1, \latoydue-.1) node[scale=.8] {${\cal O}_2$};
   \draw(\latoxdue+.1, \latoyuno+.1) node[scale=.8] {${\cal O}_2$};
   \draw(\latoxdue+.1, \latoydue-.1) node[scale=.8] {${\cal O}_2$};

   \draw (\latoxdue+2.3,-0.1) node[scale=1] {${\displaystyle= g_{12}^2 g_{34}^2 \sum_{\ulmb} M_{1,4,[\lambda_1,\lambda_2]} B_{4,\ulmb}^{(m,n)}}$};

\end{tikzpicture}
\\[.2cm]
\!\!\!\begin{tikzpicture}[scale=1.2]  
\def\shift{.35}

\def\latoxuno{-.35}
\def\latoxdue{-.38+0.8}
\def\latoyuno{.3}
\def\latoydue{-.25}

\draw (\latoxuno, \latoydue-.03) --  (\latoxdue, \latoydue-.03);
\draw (\latoxuno, \latoydue+.03) --  (\latoxdue, \latoydue+.03);
\draw (\latoxuno, \latoyuno-.03) --  (\latoxdue, \latoyuno-.03);
\draw (\latoxuno, \latoyuno+.03) --  (\latoxdue, \latoyuno+.03);

\draw[fill=white!60,draw=white] (\latoxuno-.075, \latoyuno-.075) rectangle (\latoxuno+.075, \latoyuno+.075);
\draw[fill=white!60,draw=white] (\latoxuno-.075, \latoydue-.075) rectangle (\latoxuno+.075, \latoydue+.075);
\draw[fill=white!60,draw=white] (\latoxdue-.075, \latoyuno-.075) rectangle (\latoxdue+.075, \latoyuno+.075);
\draw[fill=white!60,draw=white] (\latoxdue-.075, \latoydue-.075) rectangle (\latoxdue+.075, \latoydue+.075);

   \draw(\latoxuno-.1, \latoyuno+.1) node[scale=.8] {${\cal O}_2$};
   \draw(\latoxuno-.1, \latoydue-.1) node[scale=.8]  {${\cal O}_2$};
   \draw(\latoxdue+.1, \latoyuno+.1) node[scale=.8] {${\cal O}_2$};
   \draw(\latoxdue+.1, \latoydue-.1) node[scale=.8]  {${\cal O}_2$};

   \draw (\latoxdue+2.3,-0.1) node[scale=1] {${\displaystyle= g_{12}^2 g_{34}^2 \sum_{\ulmb} M_{2,4,[\lambda_1,\lambda_2]} B_{4,\ulmb}^{(m,n)}}$};

\end{tikzpicture}
\end{array}
\end{equation}
and our formula gives 
\begin{align}
&
\!\!\!M_{0,4,[\lambda_1,\lambda_2]}=M_{2,4,[\lambda_1,\lambda_2]}(-1)^{\lambda_1+\lambda_2} \label{A2222_fin} \\[.2cm]
&
\!\!\!M_{1,4,[\lambda_1,\lambda_2]}=M_{1,2,[\lambda_1+1]}M_{1,2,[\lambda_2]}\big((-1)^{\lambda_1}+(-1)^{\lambda_2}\big), \notag \\[.2cm]
& 
\!\!\!M_{2,4,[\lambda_1,\lambda_2]}=
M_{1,2,[\lambda_1+1]}M_{1,2,[\lambda_2]}\,
(\lambda_1{-}\lambda_2{+}1)(\lambda_1{+}\lambda_2{+}2)\,. \notag
\end{align}
In $M_{1,4}$ the sum $S_k \subseteq S_\beta$ in \eqref{Acoeffs} is non trivial. 
As the charge $p$ of the operators ${\cal O}_p$  increases, 
graphs with increasing numbers of terms in this sum will appear. 
Each term will come with a sign and a fully factorized analytic 
function of $\ulmb$. Then, by the theory independence of the $M$s, 
we can use our formula in different theories. For example, in
the ${\cal N}=0$ theory of free scalars the quantum 
numbers of the primary operators, 
dimension $\Delta$ and the spin $l$, are
$\Delta = \lambda_1+\lambda_2+\gamma$ and $l =\lambda_1-\lambda_2$ and
our (\ref{A2222_ini}-\ref{A2222_fin}) perfectly match \cite[(6.25)]{Dolan:2000ut}
for ${\cal O}_2=\vec{\Phi}.\vec{\Phi}$.
%

%==========================================================================
\section{Multiplet recombination\\ at strong coupling}
%==========================================================================

We now give a non trivial application in $\cN=4$ SYM with $SU(N)$ gauge group at large $N$.
We will be interested in extracting information 
about correlators at strong 't Hooft coupling ($g \rightarrow \infty$)  
from the free theory correlators ($g=0$), thus making use of our formula \eqref{Acoeffs}. 
In preparation for this, we recall a few important facts.
Operators appear in either $\frac{1}{2}$-BPS, 
semi-short \cite{footnote2} or long supermultiplets.   
The semi-short operators have Young diagrams 
$\ulmb=[l+2,1^a]$ and $\gamma=2a+b+2$, 
which correspond to $su(4)$ Dynkin labels $[a,b,a]$, 
twist $2a+b+2(=\gamma)$ and spin $l$. 
In the free theory, semi-short operators appear in combination 
with long operators at the superconformal unitarity bound,
because such long operators decompose
into a number of short operators when their 
anomalous dimension vanishes, i.e.~in the free theory.
Therefore, we would find generically that
   \begin{equation}\label{A=S+L}
       A_{\gamma,\ulmb}\Big|_{\rm free}= \cS_{\gamma,\ulmb} + \cK_{\gamma,\ulmb}\ \qquad \begin{array}{c} \gamma=2a+b+2 \\[.2cm] \ulmb=[l+2,1^a] \end{array}
   \end{equation}
where $\cS$ stands for semishorts, and $\cK$ for longs.
In the AdS/CFT context, the short multiplets all correspond 
to %multi-particle 
supergravity states, 
whereas the long multiplets in \eqref{A=S+L} correspond to string states. 
In the  strong coupling limit, where the theory 
is dual to AdS$_5\times$S$^5$ supergravity,
all string states gain infinite mass and disappear 
from the supergravity spectrum, therefore
 \begin{align}\label{A=S}
           A_{\gamma;\ulmb}\Big|_{\rm sugra}= \cS_{\gamma;\ulmb}\,.
   \end{align}
Since superconformal symmetry and partial 
non-renormalisation~\cite{Eden:2000bk} dictates 
that the correlator takes the form 
of the free correlator + a dynamical function, 
eq.~\eqref{A=S}  takes the form of a cancellation 
 \begin{align}\label{A=S+K-K}
           A_{\gamma;\ulmb}\Big|_{\rm sugra}= \overbrace{(\cS_{\gamma;\ulmb}+\cK_{\gamma;\ulmb})}^{\text{free}}+\overbrace{ (-\cK_{\gamma;\ulmb})}^{\text{dynamic}}\,,
   \end{align}
where the $-\cK_{}$ contribution comes from dynamics.
   
The key now is that the three-point functions of 
two half-BPS operators with a short multiplet, hence $ \cS$, 
are non-renormalised~\cite{Arutyunov:2001qw,Heslop:2001dr,Heslop:2001gp}, 
thus if we can identify them in the free theory, 
we get a prediction at strong coupling! 
These three-point couplings  are non trivial in spin 
and it is not immediate to compute 
them directly. Rather we will use another strategy, 
helped by a further non perturbative consequence 
of superconformal symmetry: three-point couplings 
of half-BPS operators $\cO_p,\cO_q$, with long 
operators of twist $p+q$ at the unitary 
bound vanish at any $g$~\cite{Arutyunov:2001qw,Heslop:2001gp}.
It follows that stringy states cannot contribute 
at $\gamma_{\rm max}$ in a correlator, and hence $\forall\,g$
\begin{align}\label{A=Sallg}
    A_{\gamma=\gamma_{\text{max}}[l+2,1^a]}^{p_1p_2p_3p_4 }(g) = \cS_{\gamma=\gamma_{\text{max}};[l+2,1^a]}^{p_1p_2p_3p_4}\,. 
    %\qquad \quad \forall\,g \ .
\end{align}
Therefore, the protected contribution $\cS^{\vec{p}}$ 
at $\gamma_{\rm max}$ is computed directly by $A^{\vec{p}}$ in the free theory.
%
%.
%
It turns out that we can do more. If there is only one short operator 
$\cO_{{\cal S};\gamma,\ulmb}$ with $\gamma=p+q$,
then we can compute its contribution to the four point correlator 
$\langle {\cO}_{p_1}{\cO}_{p_2}{\cO}_{p_3}{\cO}_{p_4}\rangle$ 
with $p_1{+}p_2{>}p{+}q$ and $p_3{+}p_4{>}p{+}q$, 
as follows:
\begin{equation}\label{SAAA1}
\!\cS^{p_1p_2p_3p_4}_{\gamma=p+q;\ulmb} =
C_{p_1 p_2\cO_{S}}
C_{p_3p_4\cO_{S}}
=\frac{A_{\gamma=p+q;\ulmb}^{p_1p_2pq}\,A_{\gamma=p+q;\ulmb}^{pqp_3p_4}}{A_{\gamma=p+q;\ulmb}^{pqpq}} 
\end{equation}
Note that 
$p+q=\gamma_{\max}$ for 
all three correlators on the RHS, since the external charges are 
$\langle p_1p_2pq\rangle$, 
$\langle pqp_3p_4\rangle$ and $\langle pqpq\rangle$.  
We thus obtain $\cS_{\gamma}^{p_1p_2p_3p_4}$ 
with $\gamma <\gamma_{\text{max}}$ 
directly from free theory data of different correlators 
at $\gamma =\gamma_{\text{max}}$.  If there is more than one short operator, 
a generalization of \eqref{SAAA1} can be  derived from a Gram matrix, as long 
as we have a sufficient number of contributing  
four-point functions to give us enough  information~\cite{Doobary:2015gia,Aprile:2019rep}.

The cancellation in \eqref{A=S+K-K} and the results 
(\ref{A=Sallg}-\ref{SAAA1}) provide a mechanism
to test results for holographic correlators in AdS$_5\times$S$^5$ supergravity. 
However, in a correlator with four single particle insertions \cite{footnote3}
eq.~\eqref{A=S+K-K} is trivial at order $O(1/N^2)$ 
because the ${\cal S}$ contribution is $O(1/N^4)$. 
The latter is in fact a piece of information in the one-loop bootstrap \cite{Aprile:2019rep}.  
Instead, when we consider a half-BPS double particle, 
$[{\cal O}_{p_1}\!{\cal O}_{p_2}]$, we find the following leading order in large $N$ behaviour:
\begin{align}
\!\!C_{[\cO_{p_1}\!\cO_{p_2}]\,\cO_q\,\cO_{\gamma<p_1+p_2+q;\ulmb}} & \sim O(1/N) \\ %\quad,\quad
C_{[\cO_{p}\cO_{q_1}][\cO_{p}\cO_{q_2}]\,\cO_{\gamma=q_1+q_2;\ulmb}} & \sim O(1)
\end{align}
and so in correlators with two single and two double-particle insertions, 
 stringy and short operators mix already at tree-level  $O(1/N^2)$.
We shall see now how this works for the first 
infinite family of such correlators.

%==========================================
\paragraph{\bf New: $\langle {\cal O}^2_2 {\cal O}^2_2 {\cal O}_q {\cal O}_q\rangle$ with $q\ge 3$.}
%==========================================
The order $O(1/N^2)$ contribution to these correlators is disconnected 
and given in terms of the known $\langle {\cO}_2{\cO}_2{\cO}_q{\cO}_q\rangle$ \cite{Uruchurtu:2008kp}. 
For short,
\begin{equation}\label{Hddqq} 
\langle{2^22^2qq}\rangle|_{\frac1{N^2}} = 2 g_{12}^2 
\langle{22qq}\rangle|_{\frac1{N^2}} \qquad  g=0\ \text{and}\ g\rightarrow \infty\,.
\end{equation}
In free theory the relation above follows from Wick's theorem. 
At strong coupling it comes from examining Witten diagrams. 
In both cases the diagrams contributing to 
$\langle{2^22^2qq}\rangle$ are the same as those of $\langle{22qq}\rangle$  
but with two extra propagators between the $\cO_2$'s.
Then \eqref{Hddqq} implies that the block decomposition 
of $\langle 2^2 2^2 q q\rangle$ mimics in every way 
that of $\langle 2 2 q q\rangle$, namely $\forall\ulmb$
\begin{align}\label{A=A}
A^{2^22^2qq}_{\gamma;\ulmb}|_{\frac1{N^2}}=2A^{22qq}_{\gamma;\ulmb}|_{\frac1{N^2}}\qquad  g=0\ \text{and}\ g\rightarrow \infty\,,
\end{align}
However in $A^{2^22^2qq}$ we expect 
semishort contributions up to twist $\gamma_\text{max}=\min(2q,8)$,
but in \eqref{A=A} only twist two and twist four  contribute, 
as for  $\langle 22qq\rangle$. Since the twist four 
contribution to $\langle 22qq\rangle$ is identical at 
strong and weak coupling, \eqref{A=Sallg}, with no stringy 
states contribution, the same must be true for 
$\langle 2^22^2qq\rangle$ at $O(1/N^2)$ even 
though $4<\gamma_{\rm max}$ for the latter.
This nicely follows from~\eqref{SAAA1}, 
by noting that ${\cal O}_2\partial^l{\cal O}_2$ is the only 
semishort double particle, and using that
\begin{equation}\label{prediction}
A^{2^22^2\!qq}_{\gamma=4}|_{g\rightarrow\infty}={\cal S}^{2^22^2\!qq}_{\gamma=4} = \frac{  A^{2^22^222}_{\gamma=4}  {A}^{22 qq}_{\gamma=4} }{  {A}^{2222}_{\gamma=4} }\Big|_{\frac{1}{N^2}}\,,
%,
\end{equation}
where  $A^{2^22^222}_{\gamma=4}=2  A^{22 22}_{\gamma=4}|_{g\rightarrow\infty}$ at $O(1)$ \cite{Aprile:2024lwy,Bissi:2024tqf}.
 Similarly, the absence of twist two stringy states at 
 $g\rightarrow \infty$ follows from the same property for $\langle 22qq\rangle$ and \eqref{A=A}.
 
We now repeat the CFT analysis for the other orientation of the correlator,
\begin{equation}\label{otherorientation2q2q}
\langle{2^2q2^2q}\rangle|_{\frac1{N^2}} = 2g_{13}^2 
\langle{2q2q}\rangle|_{\frac1{N^2}} \qquad  g=0\ \text{and}\ g\rightarrow \infty\,.
\end{equation}
In this orientation 
we expect semi-short contributions 
up to $\gamma\leq q+4=\gamma_\text{max}$ but at this order we only find
 \begin{equation}\label{diagrams2q2q}
 \!\!\!\begin{array}{c}
 \!\!\!\begin{tikzpicture} [scale=1.1] 
 \def\shift{.35}

 \def\latoxuno{-.35}
 \def\latoxdue{-.38+0.8}
 \def\latoyuno{.3}
 \def\latoydue{-.25}

 \draw (\latoxuno-.55,0.05) node[scale=.8] {$\displaystyle \frac{4 q}{N^2}$};

 \draw[fill=red!60] 
 (\latoxuno+.05, \latoydue) -- (\latoxdue, \latoyuno-.05) --  
 (\latoxdue-.05, \latoyuno) -- (\latoxuno, \latoydue+.05) -- cycle;

 \draw (\latoxuno, \latoyuno) --  (\latoxdue, \latoyuno);
 \draw (\latoxuno, \latoydue) --  (\latoxdue, \latoydue);

 \draw (\latoxdue-.1, \latoydue) --  (\latoxuno, \latoyuno-.08);
 \draw (\latoxdue, \latoydue) --  (\latoxuno, \latoyuno);
 \draw (\latoxdue, \latoydue+.08) --  (\latoxuno+.1, \latoyuno);

 \draw[fill=white!60,draw=white] (\latoxuno-.15, \latoyuno-.1) rectangle (\latoxuno+.1, \latoyuno+.1);
 \draw[fill=white!60,draw=white] (\latoxdue-.1, \latoydue-.05) rectangle (\latoxdue+.15, \latoydue+.1);

     \draw[fill=white!60,draw=white] (\latoxdue+.02, \latoyuno+.02) circle (2.5pt);
     \draw[fill=white!60,draw=white] (\latoxuno-.02, \latoydue+.02) circle (2.5pt);

    \draw(\latoxuno-.1, \latoyuno+.08) node[scale=.8] {${\cal O}^2_2$};
    \draw(\latoxuno-.04, \latoydue-.08) node[scale=.8] {${\cal O}_q$};
    \draw(\latoxdue+.14, \latoyuno+.03) node[scale=.8] {${\cal O}_q$};
    \draw(\latoxdue+.1, \latoydue-.05) node[scale=.8] {${\cal O}^2_2$};

    \draw(\latoxuno+.4, \latoydue-.5) node[scale=.7] {$\gamma=q+4$};
   
 \end{tikzpicture}
 %
 %
 %&
 %
 %
 \!\!\!\!\!\begin{tikzpicture}[scale=1.1]  
 \def\shift{.35}

 \def\latoxuno{-.35}
 \def\latoxdue{-.38+0.8}
 \def\latoyuno{.3}
 \def\latoydue{-.25}

 \draw (\latoxuno-.7,0.05) node[scale=.8] {$\displaystyle \ \ +\ \frac{4 q}{N^2}\ \ $};

 \draw[fill=red!60] 
 (\latoxuno+.05, \latoydue) -- (\latoxdue, \latoyuno-.05) --  
 (\latoxdue-.05, \latoyuno) -- (\latoxuno, \latoydue+.05) -- cycle;

 \draw (\latoxuno, \latoyuno) --  (\latoxuno, \latoydue);
 \draw (\latoxdue, \latoyuno) --  (\latoxdue, \latoydue);

 \draw (\latoxdue-.1, \latoydue) --  (\latoxuno, \latoyuno-.08);
 \draw (\latoxdue, \latoydue) --  (\latoxuno, \latoyuno);
 \draw (\latoxdue, \latoydue+.08) --  (\latoxuno+.1, \latoyuno);

 \draw[fill=white!60,draw=white] (\latoxuno-.15, \latoyuno-.1) rectangle (\latoxuno+.1, \latoyuno+.1);
 \draw[fill=white!60,draw=white] (\latoxdue-.1, \latoydue-.05) rectangle (\latoxdue+.15, \latoydue+.1);

     \draw[fill=white!60,draw=white] (\latoxdue+.02, \latoyuno+.02) circle (2.5pt);
     \draw[fill=white!60,draw=white] (\latoxuno-.02, \latoydue+.02) circle (2.5pt);

    \draw(\latoxuno-.1, \latoyuno+.08) node[scale=.8] {${\cal O}^2_2$};
    \draw(\latoxuno-.04, \latoydue-.08) node[scale=.8] {${\cal O}_q$};
    \draw(\latoxdue+.14, \latoyuno+.03) node[scale=.8] {${\cal O}_q$};
    \draw(\latoxdue+.1, \latoydue-.05) node[scale=.8] {${\cal O}^2_2$};

    \draw(\latoxuno+.4, \latoydue-.5) node[scale=.7] {$\gamma=q+2$};
 
 \end{tikzpicture}
 \!\!\!\begin{tikzpicture}[scale=1.1] 
 \def\shift{.35}

 \def\latoxuno{-.35}
 \def\latoxdue{-.38+0.8}
 \def\latoyuno{.3}
 \def\latoydue{-.25}

 \draw (\latoxuno-1,0.05) node[scale=.8] {$\displaystyle\ +\ \frac{4 q(q-1)}{N^2}\ $};

 \draw[fill=red!60] 
 (\latoxuno+.05, \latoydue) -- (\latoxdue, \latoyuno-.05) --  
 (\latoxdue-.05, \latoyuno) -- (\latoxuno, \latoydue+.05) -- cycle;

 \draw (\latoxuno, \latoyuno) --  (\latoxuno, \latoydue);
 \draw (\latoxdue, \latoyuno) --  (\latoxdue, \latoydue);
 \draw (\latoxuno, \latoydue) --  (\latoxdue, \latoydue);
 \draw (\latoxuno, \latoyuno) --  (\latoxdue, \latoyuno);

 \draw (\latoxdue-.05, \latoydue) --  (\latoxuno, \latoyuno-.03);
 \draw (\latoxdue, \latoydue+.03) --  (\latoxuno+.05, \latoyuno);

 \draw[fill=white!60,draw=white] (\latoxuno-.15, \latoyuno-.1) rectangle (\latoxuno+.1, \latoyuno+.1);
 \draw[fill=white!60,draw=white] (\latoxdue-.1, \latoydue-.05) rectangle (\latoxdue+.15, \latoydue+.1);

     \draw[fill=white!60,draw=white] (\latoxdue+.02, \latoyuno+.02) circle (2.5pt);
     \draw[fill=white!60,draw=white] (\latoxuno-.02, \latoydue+.02) circle (2.5pt);

    \draw(\latoxuno-.1, \latoyuno+.08) node[scale=.8] {${\cal O}^2_2$};
    \draw(\latoxuno-.04, \latoydue-.08) node[scale=.8] {${\cal O}_q$};
    \draw(\latoxdue+.14, \latoyuno+.03) node[scale=.8] {${\cal O}_q$};
    \draw(\latoxdue+.1, \latoydue-.05) node[scale=.8] {${\cal O}^2_2$};

    \draw(\latoxuno+.4, \latoydue-.5) node[scale=.7] {$\gamma=q+2$};

 \end{tikzpicture}
 \end{array}\vspace{-3mm}
 \end{equation}
Differently from $\langle 2^2 2^2 qq\rangle$, 
the relation \eqref{otherorientation2q2q} no longer implies a 
 relation between the block coefficients of $\langle 2^2 q2^2q\rangle$ 
 and those of $\langle 2q2q\rangle$, 
because the additional $g_{13}^2$ changes the value 
of $\gamma$, see \eqref{defgammak}, and so the blocks. 
Therefore we expect to find new stringy contributions 
compared to  $\langle{2q2q}\rangle$, precisely at $\gamma=q+2$, 
and consequently a new type of cancellation \eqref{A=S+K-K} at $g\rightarrow\infty$.
Although there are many short double particles at twist $q+2$ only 
${\cal O}_2\partial^l {\cal O}_{q}$ contributes in \eqref{diagrams2q2q}, thus
\begin{equation}\label{otherprediction}
{\cal S}^{2^2q2^2 q}_{\gamma=q+2} \Big|_{\frac{1}{N^2}}= \frac{  {\cal S}^{2^2q 2q}_{\gamma=q+2} \times {\cal S}^{ 2q2^2q}_{\gamma=q+2} }{  {\cal S}^{2q2q}_{\gamma=q+2} }\Big|_{\frac{1}{N^2}}\,.
\end{equation}
On the RHS of \eqref{otherprediction} the leading contributions come from 
the \emph{disconnected} diagrams drawn below \cite{footnote4}
\begin{equation} 
\begin{array}{ccc}
\begin{tikzpicture}[scale=1]  
\def\shift{.35}

\def\latoxuno{-.35}
\def\latoxdue{-.38+0.8}
\def\latoyuno{.3}
\def\latoydue{-.25}

\draw[fill=red!60] 
(\latoxuno+.05, \latoydue) -- (\latoxdue, \latoyuno-.05) --  
(\latoxdue-.05, \latoyuno) -- (\latoxuno, \latoydue+.05) -- cycle;

\draw (\latoxdue-.05, \latoydue) --  (\latoxuno, \latoyuno-.05);
\draw (\latoxdue, \latoydue+.05) --  (\latoxuno+.05, \latoyuno);

\draw[fill=white!60,draw=white] (\latoxuno-.15, \latoyuno-.1) rectangle (\latoxuno+.1, \latoyuno+.1);
\draw[fill=white!60,draw=white] (\latoxdue-.1, \latoydue-.05) rectangle (\latoxdue+.15, \latoydue+.1);

    \draw[fill=white!60,draw=white] (\latoxdue+.02, \latoyuno+.02) circle (2.5pt);
    \draw[fill=white!60,draw=white] (\latoxuno-.02, \latoydue+.02) circle (2.5pt);

   \draw(\latoxuno-.1, \latoyuno+.08) node[scale=.8] {${\cal O}_2$};
   \draw(\latoxuno-.04, \latoydue-.08) node[scale=.8] {${\cal O}_q$};
   \draw(\latoxdue+.14, \latoyuno+.03) node[scale=.8] {${\cal O}_q$};
   \draw(\latoxdue+.1, \latoydue-.05) node[scale=.8] {${\cal O}_2$};

   \draw(\latoxdue+1.6, \latoydue+.2) node[scale=1] {$;$};
 
\end{tikzpicture} 
&
\qquad\qquad
&
\!\!\!\!\!\begin{tikzpicture}[scale=1]  
\def\shift{.35}

\def\latoxuno{-.35}
\def\latoxdue{-.38+0.8}
\def\latoyuno{.3}
\def\latoydue{-.25}

\draw[fill=red!60] 
(\latoxuno+.05, \latoydue) -- (\latoxdue, \latoyuno-.05) --  
(\latoxdue-.05, \latoyuno) -- (\latoxuno, \latoydue+.05) -- cycle;

\draw (\latoxuno-.1, \latoyuno) --  (\latoxuno-.1, \latoydue+.1);
\draw (\latoxuno-.1, \latoyuno) --  (\latoxdue, \latoyuno);

\draw (\latoxdue-.05, \latoydue) --  (\latoxuno, \latoyuno+.2-.05);
\draw (\latoxdue, \latoydue+.05) --  (\latoxuno+.05, \latoyuno+.2);

\draw[fill=white!60,draw=white] (\latoxuno-.2, \latoyuno-.1) rectangle (\latoxuno, \latoyuno+.1);
\draw[fill=white!60,draw=white] (\latoxdue-.1, \latoydue-.05) rectangle (\latoxdue+.15, \latoydue+.1);

    \draw[fill=white!60,draw=white] (\latoxdue+.02, \latoyuno+.02) circle (2.5pt);
    \draw[fill=white!60,draw=white] (\latoxuno-.02, \latoydue+.02) circle (2.5pt);

\draw(\latoxuno-.1, \latoyuno+.2+.08) node[scale=.8] {${\cal O}_2$};
   \draw(\latoxuno-.2, \latoyuno+.08) node[scale=.8] {${\cal O}_2$};
   \draw(\latoxuno-.04, \latoydue-.08) node[scale=.8] {${\cal O}_q$};
   \draw(\latoxdue+.14, \latoyuno+.03) node[scale=.8] {${\cal O}_q$};
   \draw(\latoxdue+.1, \latoydue-.05) node[scale=.8] {${\cal O}_2$};

 \draw(\latoxdue+.6, \latoydue+.2) node[scale=1] {$.$};
 
\end{tikzpicture} 
\end{array}
 \end{equation}
Thus ${\cal S}^{2^2q 2q}_{\gamma=q+2}=2qM^{4q2q}_{\gamma=q+2}$ 
and ${\cal S}^{2q2q}_{\gamma=q+2}=M^{2q2q}_{\gamma=q+2}$ in \eqref{otherprediction}.
Now, from \eqref{A=S+K-K} we read off the stringy 
contribution to recombine, e.g.~in the rep $[0,q,0]$ we find
\begin{equation}\label{K}
\!\!\!\!\!\!\!\!{\cal K}^{2^2\!q2^2\!q}_{[0q0]}\Big|_{\frac{1}{N^2}}\!\!\!\!= - {\cal S}^{2^2\!q2^2\!q}_{\gamma=q+2,[2+l]}\Big|_{\frac{1}{N^2}}
+A^{2^2\!q2^2\!q}_{\gamma=q+2,[2+l]}\Big|_{\frac{1}{N^2}}
\end{equation}
where $A^{2^2\!q2^2\!q}_{q+2,[2+l]}=\frac{4q}{N^2} \left[ M^{4q4q}_{0,q+2,[2+l]}\!+\!(q-1)M^{4q4q}_{1,q+2,[2+l]}\right]$.\vspace{2mm}

Beautifully, the non trivial spin prediction in \eqref{K} is canceled by the dynamical 
part, which from \eqref{otherorientation2q2q} reads
\begin{equation}\label{Hdqdq} 
%{\cal H}_{2^2q2^2q} =  
g_{13}^4 g_{24}^q\, {\cal I}(x_i,y_j)\bigg[
-\frac{4q}{N^2}\frac{(y_1 y_2)^2 \overline{D}_{2,q+2,2,q}(x_1,x_2)}{(q-2)!} 
 \bigg] 
\end{equation}
with ${\cal I}=\prod_{i,j=1}^{2}(x_i-y_j)/(y_1 y_2)^2$. One can derive further 
predictions at twist $\gamma=q+2$ for different $su(4)$ reps, 
 and all  are consistent with~\eqref{Hdqdq}. 
 To illustrate the non trivial nature of these agreements, 
 note that~\eqref{Hdqdq} is given by a single $\overline{D}$ function,  
 which then suggests that the predicted $\cK$s 
 must be related to each other. Indeed we find
\begin{equation}
\frac{q-3}{q-1}{\cal K}^{2^2\!q2^2\!q}_{[2,q-4,2]} =\frac{-q}{2(q-2)} {\cal K}^{2^2\!q2^2\!q}_{[1,q-2,1]}={\cal K}^{2^2\!q2^2\!q}_{[0,q,0]}\Big|_{\frac{1}{N^2}}\,.
\end{equation}
The precise knowledge of ${\cal S}$ is crucial in each case.
Additional details are given in the supplementary material, ,
where we also consider  examples with more general double 
particle external operators, such as $[\cO_3^2]$ and $[\cO_2\cO_4]$. 
All the relevant formulae are immediately \emph{writable and computable}
in terms of the $M$ coefficients, 
and this reduces dramatically the CFT analysis.

\section{\bf Outlook}
We obtained an explicit formula for the block 
expansion of \emph{all} free theory scalar diagrams in 1d, 2d, and 4 dimensional CFTs 
with ${\cal N}=0,2,4$ SUSYs. Our formula applies to (generalized) free theories 
both in perturbation theory and at strong coupling, e.g.~in holographic CFTs.  
 We note here that our approach can be extended to spinor 
and vector propagators. In this case a free diagram is reconstructed 
by considering more general monomials in the $x_i,y_j$ variables.
These can be accommodated in a more general
Cauchy identity \cite{Aprile:2021pwd} with superJacobi polynomials on the RHS.
In the supplemental material 
we give a \emph{new} determinant formula for them showing that the 
$M_{G,\gamma,\ulmb}$ are evaluation formulas of determinants 
\emph{for all free theories} in 4d, 2d, 1d. 

Free and generalized free theories occupy a special place in the conformal bootstrap 
and we believe that our formula will be useful 
in this context. As a first indication of this we illustrated the 
utility of our formula in computing new data 
for the mixed correlation functions with 
two single- and two double-particle insertions 
in ${\cal N}=4$ SYM at large $N$ 
and strong 't Hooft coupling. Notably, this
requires an improved understanding of the recombination 
of stringy states at the unitarity boundy, which 
accidentally was trivial for single particle correlators. 
Our results give new dynamical information at strong coupling, 
that we tested successfully in $\langle {\cO}_2^2{\cO}_2^2{\cO}_q{\cO}_q\rangle$. 
Our method applies more generally, and therefore it would be interesting to understand 
the relation with the $10d$ conformal symmetry \cite{Caron-Huot:2018kta,Abl:2020dbx}, 
since it is tempting to conjecture that there exist generating functions such 
as $\langle\phi^2\phi\phi\phi\rangle$, $\langle\phi^2\phi^2\phi\phi\rangle\sim\langle \phi\phi\rangle\langle\phi\phi\phi\phi\rangle$, 
and multiparticle generalizations thereof, that compute  
tree-level/disconnected correlators for all the AdS$_5\times$S$^5$ mixed correlators.

\section{ACKNOWLEDGEMENTS}

We thank Rodolfo Russo and Stefano Giusto for useful discussions and comments.
F. Aprile is supported by RYC2021-031627-I funded
by MCIN/AEI/10.13039/501100011033 and by the Eu-
ropean Union NextGenerationEU/PRTR, and by the MSCA 
programme - HeI - \hyperlink{https://cordis.europa.eu/project/id/101182937}{DOI 10.3030/101182937}.
M. Santagata is supported by the Ministry of Science and Technology (MOST) through the grant 110-
2112-M-002-006-. P. Heslop is supported in part by STFC under grant
ST/X000591/1.

\resumetoc
\onecolumngrid

\vspace{2cm}

\setcounter{equation}{0} % Reset equation counter
\renewcommand{\theequation}{S\arabic{equation}} % Redefine equation numbering
\setcounter{secnumdepth}{2}

\begin{center}
\textbf{\large Supplemental Materials}
\end{center}

\tableofcontents

%=========================================================================================
\section{From Young diagrams to superconformal representations}
%=========================================================================================
\label{sec:I}
The external operators considered in the main letter are scalar fields of a Grassmannian field theory \cite{Doobary:2015gia}. 
By this we mean a theory with $SU(m, m|2n)$ symmetry living on the (complexified) 
space $Gr(m|n, 2m|2n)$  of $m|n$-planes in $2m|2n$ dimensions. 
In this language, Minkowski space corresponds to $(m,n)=(2,0)$, and the corresponding field theory is an ordinary 
4d non supersymmetric theory whose fundamental fields $\Phi$ are scalars. 
In the case of arbitrary $m$ we will think of these theories as non 
supersymmetric generalized free theories. Then, $\cN=4$ SYM corresponds to 
$(m,n)=(2,2)$, and the fundamental field $\Phi$ is the field strength multiplet. Finally, 
4d $\cN=2$ theories correspond to $(m,n)=(2,1)$ where $\Phi$ is the hypermultiplet.\\ 

In any Grassmannian field theory the fields exchanged in the OPE between scalar 
operators are specified by $\gamma$ and a Young diagram $\ulmb$ that must fit into a 
fat hook shape with at most $m$ long rows and $n$ long columns. More precisely: 
if we specify the Young diagram by its row lengths, $\ulmb=[\lambda_1,\lambda_2,..]$  this condition implies 
$\lambda_{i>m} \leq n$. Similarly, if we specify $\ulmb$ by its column lengths, $\lambda'_i$, we have 
$\lambda_{i>n}' \leq m$. For example, in the $(m,n)=(2,0)$ theory $\ulmb$ has at most 
two rows $\ulmb=[\lambda_1,\lambda_2]$, and the quantum numbers $\tau=\Delta{-}l$ and $l$, 
respectively, the twist and the spin, are  given by
\begin{equation}\label{two_rows_QN}
\tau = 2\lambda_2+\gamma\qquad;\qquad l=\lambda_1-\lambda_2\,.
\end{equation}
Note that we have three quantum numbers $\gamma,\lambda_1,\lambda_2$ 
rather than two $\Delta,l$, and indeed $\Delta,l$ are invariant under 
$\lambda_i\rightarrow\lambda_i+1$, $\gamma\rightarrow \gamma-2$. The additional 
quantum number is the number operator $\gamma$, counting the number of fundamental 
fields constituting the operator and which is preserved in free theory but not in the interacting theory. 
As  a simple illustrative example, consider $\Phi^4$ and $\Phi \Box \Phi$ 
both have $\Delta=4, l=0$, but $\Phi^4$ has $\gamma=4,\lambda_i=0$ whereas 
$\Phi \Box \Phi$ has $\gamma=2,\lambda_i=1$. These operators can be distinguished 
in the free theory, but mix in the interacting theory.\\

In supersymmetric theories, $n,m\neq0$. We say that $\ulmb$ is a typical 
representation if $m^n\subseteq \ulmb$, otherwise we call it atypical. 
Atypical Young diagrams correspond to operators in short supermultiplets 
whereas typical diagrams label long multiplets.\\

In the case of $\mathcal{N}=4$ SYM, (i.e. $m=2=n$) the values of $\gamma,\ulmb$ 
are related to the twist, $\Delta-l$, the spin, $l$, and the internal $su(4)$ rep, $[aba]$, 
of the exchanged primary as given in the table below
\begin{equation}\label{tableQN1}
\begin{array}{|c|c|c|c|c|}
%& 
\hline
\textrm{Multiplet} & \ulmb & \tau &  l & su(4) \\[.3ex]
\hline
\hline
\frac{1}{2}\textrm{-BPS}  & [\varnothing] &\rule{0pt}{.4cm}\gamma & 0 & [0,\gamma,0] \\[1ex]
\frac{1}{4}\textrm{-BPS}  &  [1^{\mu}] &\gamma & 0 & [\mu,\gamma-2\mu,\mu]\\[1ex]
\textrm{semi-short}
&  [\lambda,1^{\mu}] &\gamma & \lambda-2& [\mu,\gamma-2\mu-2,\mu]\\[1ex]
\hline
\end{array}
\end{equation}
and for long multiplets $\ulmb_{}= 
[\lambda_1,\lambda_2,2^{\mu_2},1^{\mu_1}]$ corresponds to
\begin{equation}\label{tableQN2}
\tau=\gamma+2\lambda_2-4\qquad;\qquad l=\lambda_1-\lambda_2
\end{equation}
with $su(4)$ labels 
$[\mu_1,\gamma-2\mu_1-2\mu_2-4,\mu_1]$.
In the free theory $\tau_{\rm\, long}\ge 2a + b + 4$. 
As for the non supersymmetric case, there is again an ambiguity
$\lambda_i\rightarrow \lambda_i+1$, $\mu_2\rightarrow \mu_2-1,\gamma\rightarrow \gamma-2$ 
for long multiplets, reflecting mixing of operators with differing numbers of constituent fundamental fields, $\gamma$.\\

For $\mathcal{N}=2$ SCFTs we refer directly to \cite{Doobary:2015gia}.\\

It will be useful to make contact between the Young diagram notation, that we use to 
label the (super-)primary operators and the block coefficients $A$ (as well as the $M$aster coefficients), 
and the ${\cal N}=4$ notation that is often used in the literature. We will do this in the very well known 
example of the four-point function of the chiral primary operator in the stress tensor multiplet,
\begin{equation}\label{repet2222}
\langle {\cal O}_2 {\cal O}_2 {\cal O}_2 {\cal O}_2 \rangle=4a^2 g_{12}^2 g_{34}^2\Bigg[ 1+U^2\sigma^2 
+\frac{U^2\tau^2}{V^2}+ \frac{4}{a}\Bigg[ U\sigma +\frac{U\tau}{V} + \frac{U^2\sigma\tau}{V}\Bigg] \Bigg] 
\quad;\quad U\sigma=\frac{ g_{13} g_{24} }{ g_{12} g_{34} }\quad;\quad \frac{\tau}{\sigma V}= 
\frac{g_{14} g_{23} }{ g_{13} g_{24} }
\end{equation}
Table \ref{table1} summarises the comparison of our block expansion (in the interacting theory  
with twist-two semishort operators recombined) with results from the 
literature \cite[(3.14)]{Dolan:2004iy}, and we find perfect agreement.
\begin{figure*}[!htbp]
\def\arraystretch{1.5}

\begin{tabular}{|c|c|c|}
\hline
 Multiplet &  \cite[(3.14)]{Dolan:2004iy} &  external charges $\vec{p}=2222$    \\[0.5ex]
\hline\hline
${\cal B}_{[000]}$  & $1$  & $M_{0,0,[\varnothing]}$  \\[1ex] 
\hline
${\cal B}_{[020]}$  & $6C_{11}$  & $\frac{4}{a}( M_{0,2,[\varnothing]}+M_{1,2,[\varnothing]})$  \\[1ex] 
\hline
${\cal B}_{[040]}$ & $60C_{2}$  & $M_{0,4,[\varnothing]}+M_{2,4,[\varnothing]}+\frac{4}{a}M_{1,4,[\varnothing]}$  \\[1ex]
\hline
 ${\cal B}_{[202]}$  &   $9 C_{20}$     &  
 $M_{0,4,[1,1]}+M_{2,4,[1,1]}+ \frac{4}{a} M_{1,4,[1,1]} -  \frac{4}{a} M_{0,2,[2]}- \frac{4}{a} M_{1,2,[2]}$ 
 \\[1ex]   
\hline
\hline
 ${\cal S}_{[020],l=0,2,\ldots} $     &  $2^{-l-2}A_{00,0,l+2}$   & $M_{0,4,[2+l]}+M_{2,4,[2+l]}+ \frac{4}{a}M_{1,4,[2+l]}$  \\[1ex]
 \hline
   ${\cal S}_{[101],l=1,3,\ldots}$   &  $-2^{-l-3} b_{0,l+2}$  \rule{0pt}{0cm}  & 
 $
 M_{0,4,[2+l,1]}+M_{2,4,[2+l,1]}+ \frac{4}{a} M_{1,4,[2+l,1]} - \frac{4}{a} M_{0,2,[3+l]}-\frac{4}{a} M_{1,2,[3+l]}$ \\[1ex]
\hline
\hline
${\cal L}_{[000],t=1,l=0,2,\ldots}$ & $2^{-l}\hat{A}_{00,1,l}$ &  $\frac{4}{a} M_{0,2,[2+l]}+\frac{4}{a} M_{1,2,[2+l]}$   \\[1ex]
\hline
 ${\cal L}_{[000],t\ge2,l=0,2,\ldots}$ & $2^{-l} A_{00,t,l}$ &  
 $M_{0,4,[t+l,t]}+M_{2,4,[t+l,t]}+ \frac{4}{a}M_{1,4,[t+l,t]}$ \\[1ex]
\hline 
\end{tabular}
\renewcommand{\figurename}{Table}
\caption{  
The $M$ coefficients are explicit in \eqref{A2222_ini}-\eqref{A2222_fin}. $a=N^2-1$. We used ${\cal B}$ for $\frac{1}{2}$- and $\frac{1}{4}$-BPS operators, and ${\cal S}$ for semi-short operator. In  both cases $\tau$ is fixed by the $su(4)$ rep. We used 
${\cal L}_t$ for long multiplets, where $t=\frac{\Delta-l}{2}$. 
\label{table1}}
\end{figure*}

~\\
Finally, we point out that for correlators with higher external charges, we usually have to sum over 
Young diagrams with different $\gamma$, to account for mixing of operators with different numbers of 
constituent fields,  in order to obtain the block coefficient in the long sector. For example, in 
$\langle \cO_3\cO_3\cO_3\cO_3\rangle$ the total contribution to the long sector in the $[000]$ 
rep is ${\cal L}_{[000],t= 2,l}= A_{4,[\lambda_1,2]}$ and 
${\cal L}_{[000],t\ge 3,l}=A_{6,[\lambda_1,\lambda_2,2]}+A_{4,[\lambda_1+1,\lambda_2+1]}$
where $\lambda_2=t-1\,,l=\lambda_2-\lambda_1$.

%=========================================================================================
\section{Semishort unmixing in $AdS_5\times S^5$ supergravity: explicit results}
%=========================================================================================
\label{sec:III}

In this appendix we expand the discussion about 
the semishort sector in correlators with two single and two double particle 
insertions. In the process we give explicit details about various results that were mentioned in the main text.\\
To simplify notation in various places we will write the semi-short block coefficients as
   \begin{align}
      {\cal S}^{\vec{p}}_{[a,b,a]}:=A^{\vec{p}}_{ {\cal S}, {\gamma=b+2a+2;[l+2,1^a]}}\ .
   \end{align}

\subsection{ Exact CFT data for the twist $4$ semishort sector}
The twist four semishort sector consists of the double trace operators
\begin{equation}\label{twist4sector}
{\cal O}_2\partial^l{\cal O}_2\Big|_{[020],l=0,2,\ldots}\qquad;\qquad {\cal O}_2\partial^l{\cal O}_2\Big|_{[101],l=1,3,\ldots}
\end{equation}
These are the only semishort protected operators in these reps, 
since the lightest  protected triple trace operators necessarily have twist six. \\

In the main text we showed that the twist four semishort sector contributes 
to the correlators $\langle {\cal O}_2^2 {\cal O}_2^2{\cal O}_q {\cal O}_q\rangle$ ($q\ge 3$) 
with an OPE coefficient that can be derived from the relation~\eqref{prediction} and reads
\begin{equation}\label{suppl_prediction}
 {\cal S}^{2^22^2\!qq}_{[aba]} = \frac{  {\cal S}^{2^22^222}_{[aba]} \times {\cal S}^{22 qq}_{[aba]} }{  {\cal S}^{2222}_{[aba]} }\qquad;\qquad [aba]=[020],[101]\,.
\end{equation}
We emphasize here that ${\cal S}^{2^22^2\!qq}_{[aba]}$ is an exact 
quantity in $a=N^2-1$ because, as mentioned above, there are no other semishort 
operators at twist four in the reps $[020]$ and $[101]$.
While the color factor of $\langle {\cal O}_2 {\cal O}_2{\cal O}_2 {\cal O}_2\rangle$ 
can be read from \eqref{repet2222}, it is not difficult to compute the others,  
\begin{align}
\label{mixed2d2d22}
\langle {\cal O}^2_2{\cal O}^2_2{\cal O}_2{\cal O}_2\rangle &=  {\cal N}_{[2^2]}{\cal N}_{2}\ g_{12}^4 g_{34}^2\ \Bigg[ 
1+ \left(2+\frac{4}{a}\right)\Bigg[ U^2\sigma^2 +  \frac{U^2\tau^2}{V^2} \Bigg]+ \frac{8}{a}\Bigg[ U\sigma +\frac{U\tau}{V}\Bigg] +   \frac{24}{a}\frac{ U^2\sigma\tau}{V} \Bigg]\\[.2cm]
\label{singlep22qq}
\langle {\cal O}_2{\cal O}_2{\cal O}_q{\cal O}_q\rangle & = {\cal N}_{2} {\cal N}_{q}\ g_{12}^2 g_{34}^q\ \Bigg[ 
1+ \frac{2q}{a}\Bigg[ U\sigma +\frac{U\tau}{V}\Bigg] + \frac{2 q(q-1)}{a} \frac{U^2\sigma\tau}{V} \Bigg] \qquad q>2
\end{align}
where the two-point function normalizations are
\begin{equation}
    {\cal N}_{[2^2]}=8a(a+2)\qquad;\qquad 
    {\cal N}_q=\frac{ q^2(q-1) }{ (N-q+1)^{-1}_{q-1} - (N+1)^{-1}_{q-1} }\,.
\end{equation}
As a function of $N$, the function ${\cal N}_q$ is a rational function of $N$, written in terms 
of Pochhammer symbols. In the large $N$ limit, ${\cal N}_q\simeq qN^{q}$. We will 
usually switch from $N$ to $a=N^2-1$. For example ${\cal N}_2=2a$.\\

Both \eqref{mixed2d2d22}-\eqref{singlep22qq} are next-to-next-to extremal correlators, 
which implies six possible propagator structures. The fact that only four appears 
in \eqref{singlep22qq} is because of the definition of single particle operators \cite{Aprile:2020uxk}.\\ 

Coming back to \eqref{suppl_prediction}, the ${\cal S}^{2222}$ coefficients 
are given in Table~\ref{table1}, and for the others we find 
\begin{equation}
\begin{array}{|ccl|}
\hline
\rule{0pt}{0.5cm}\vec{p} & =  & \{ 2,2,q,q\} \\[1ex]
\hline
\rule{0pt}{.6cm} {\cal S}_{[020]} & = & \frac{2q(q-1)}{a} M_{1,4,[2+l]}  \\
\rule{0pt}{.6cm} {\cal S}_{[101]} & = & \frac{2q(q-1)}{a} M_{1,4,[2+l,1]} - \frac{2q}{a} 
( M_{0,2,[3+l]} + M_{1,2,[3+l]} ) \\[1ex]
\hline\hline 
\rule{0pt}{0.45cm}\vec{p} & = & \{[2^2],[2^2],2,2\} \\[1.1ex]
\hline
\rule{0pt}{.6cm} {\cal S}_{[020]}  & = & (2+\frac{4}{a})( M_{0,4,[2+l]}+ M_{2,4,[2+l]}) + \frac{24}{a} M_{1,4,[2+l]} \\
\rule{0pt}{.6cm} {\cal S}_{[101]} & = & 
 (2+\frac{4}{a})( M_{0,4,[2+l,1]}+ M_{2,4,[2+l,1]}) + \frac{24}{a} M_{1,4,[2+l,1]} -  \frac{8}{a} 
( M_{0,2,[3+l]} + M_{1,2,[3+l]} )\\[1ex]
\hline
\end{array}
\end{equation}
All together, the non perturbative results for ${\cal S}^{2^22^2 qq}$ are 
\begin{align}
&
\!\!\!\!{\cal S}^{2^22^2 qq}_{[020]}\!=+8q(q-1) \frac{a+2}{a^2} \,\Bigg[\!\frac{ (l+3)(l+4) +\frac{12}{a+2} }{ \!\!(l+3)(l+4) +\frac{4}{a} }\!\Bigg]\, \frac{ (l+3)!^2}{(2l+6)!}\, \frac{1+(-1)^l}{2}\,,\notag
 \\
 &
 \!\!\!\!{\cal S}^{2^22^2 qq}_{[101]}\!=-4q(q+1) \frac{a+2}{a^2} \,\Bigg[\!\frac{ (l+2)(l+5) -\frac{20}{a+2} }{\!\! (l+2)(l+5) -\frac{12}{a} }\!\Bigg]\,\frac{ (l+3)!^2}{(2l+6)!}\,\frac{1-(-1)^l}{2} \,.\notag
\end{align}
As explained already in the main text, the leading $O(1/{N^2})$ contribution is given 
by disconnected diagrams, and these give  the same contribution as the disconnected 
diagrams in $\langle {\cal O}_{2^2} {\cal O}_{2^2} {\cal O}_{q} {\cal O}_{q}\rangle$.  
Indeed, from \eqref{mixed2d2d22} we find
\begin{align}
%\!\!\!\!
{\cal S}^{2^22^2 qq}_{[020]}\Big|_{\frac{1}{a}} & = 4 q(q-1) M^{44 qq}_{1,4,[2+l] }= A^{2^22^222}_{[020]}\Big|_{\frac{1}{a}} \\
{\cal S}^{2^22^2qq}_{[101]}\Big|_{\frac{1}{a}} &= 4 q(q-1) M^{44 qq}_{1,4, [2+l,1] } - 4q \left(M^{44qq}_{0,4,0,[l+3]}+M^{44qq}_{1,4,[l+3]} \right)=A^{2^22^222}_{[101]}\Big|_{\frac{1}{a}}
\end{align}
It follows that in these reps no stringy contribution is present at order $\frac{1}{N^2}$.

\subsection{Exact CFT  data for the twist $5$ semishort sector}
The twist five semishort sector consists of double trace operators
\begin{equation}\label{twist5sector}
{\cal O}_2\partial^l{\cal O}_3\Big|_{[030],l=0,1,2,3\ldots}\qquad;\qquad {\cal O}_2\partial^l{\cal O}_3\Big|_{[111],l=0,1,2,3,\ldots}
\end{equation}
As for the twist four sector, these are the only semishort operators in this reps. \\

The semishorts operators \eqref{twist5sector} enter the OPE analysis of 
$\langle {\cal O}_2{\cal O}_3{\cal O}_2{\cal O}_3\rangle$ and that of higher charges correlators. For example,
the simplest mixed correlator where both twist four and twist five protected 
double particle operators are crucial are $\langle {\cal O}_2^2{\cal O}^2_2{\cal O}_3{\cal O}_3\rangle$ 
and its crossed orientation $\langle {\cal O}_3{\cal O}^2_2{\cal O}_3{\cal O}^2_2\rangle$. 
For the latter, the use of \eqref{SAAA1} immediately gives 
\begin{equation}
 {\cal S}^{32^232^2}_{[aba]}=  \frac{  {\cal S}^{32^223}_{[aba]} \times {\cal S}^{23 32^2}_{[aba]} }{  {\cal S}^{2323}_{[aba]} }\qquad [aba]=[030],[111]\,.
\end{equation}
To compute this prediction we need the free theory for $\langle \cO_2\cO_3\cO_2\cO_3\rangle$, that can be derived from \eqref{singlep22qq}, and the free theory for,
\begin{equation}
\frac{ \langle {\cal O}_2{\cal O}_3{\cal O}_3{\cal O}_2^2\rangle}{ \frac{72}{N} (N^2-1)(N^2+1) (N^2-4)} = g_{12}^3 g_{24}^{} g_{34}^3 \Bigg[  {\frac{U^2\tau}{V^2} + \frac{4}{(a+2)}\frac{U^2\sigma\tau}{V} }+ \frac{2}{(a+2)} \Bigg[ U\sigma + \frac{2U\tau}{V}\Bigg] \Bigg]
\end{equation}
Writing the prediction in terms of the $M$ coefficients is straightforward, and we find that
\begin{equation}
{\cal S}^{32^232^2}_{[030]}= 
\left\{\begin{array}{rl}   
\displaystyle
\frac{144 (l+5) \left( (l+3)+\frac{4}{a+2} \right)^{\!2} }{  (l+3)(l+4) +\frac{12}{a}  }&\!\!\!\times\ 
\displaystyle
\frac{(l+4)!}{(2l+8)!} \frac{1+(-1)^l}{2} \\[.6cm]
\displaystyle
\frac{144 (l+3) \left( (l+5)-\frac{4}{a+2} \right)^{\!2} }{  (l+4)(l+5) +\frac{12}{a} }&\!\!\!\times\ 
\displaystyle
\frac{(l+4)!}{(2l+8)!} \frac{1-(-1)^l}{2} 
\end{array} \right. 
\end{equation}
and 
\begin{equation}
{\cal S}^{32^232^2}_{[111]}= 
\left\{\begin{array}{rl}   \displaystyle
\frac{72}{3}\frac{(l+2) \left( (l+6)+\frac{10}{a+2} \right)^{\!2} }{  (l+4)(l+6) -\frac{24}{a}  }&\!\!\!\times\ 
\displaystyle
\frac{(l+4)!}{(2l+8)!} \frac{1+(-1)^l}{2} \\[.6cm]
\displaystyle
\frac{72}{3} \frac{ (l+6) \left( (l+2)-\frac{10}{a+2} \right)^{\!2} }{  (l+2)(l+4) -\frac{24}{a}  }&\!\!\!\times\ 
\displaystyle
\frac{(l+4)!}{(2l+8)!} \frac{1-(-1)^l}{2} 
\end{array} \right. 
\end{equation}
As for the twist four semishort sector, we emphasize that these results are exact in 
$a=N^2-1$ because the double trace operators \eqref{twist5sector} are the only twist five semishort operators. 

%================================================
\subsection{An example with a $3\times3$ Gram matrix} 
%================================================
In this section we discuss examples of mixed single- and double-particle correlators where 
the prediction of the double particle protected sector requires the study of a 3$\times 3$ 
Gram matrix. We shall consider the case of semishort operators belonging to the $[040]$ rep, and the correlators 
\begin{equation}
\langle {\cal O}_2^2  {\cal O}_2^2 {\cal O}_4 {\cal O}_4 \rangle\qquad;\qquad 
\langle {\cal O}_2  {\cal O}_2  [{\cal O}_4{\cal O}_2] [{\cal O}_4{\cal O}_2]\rangle 
\qquad;\qquad \langle {\cal O}_2   {\cal O}_2 {\cal O}_3^2 {\cal O}_3^2\rangle.
\end{equation}
Let us first comment on the supergravity  results for the correlators of interest. We will write them in the form
\begin{equation}
\langle {\cO}_{p_1}{\cO}_{p_2}{\cO}_{p_3}{\cO}_{p_4}\rangle = {\rm free} + {\cal P}[g_{ij}] {\cal I}(x_1,x_2,y_1,y_2) {\cal H}_{\vec{p}}(x_1,x_2,y_1,y_2)
\end{equation}
where ${\cal I}=\prod_{i,j}(x_i-y_j)/(y_1y_2)^2$.
Then for the mixed correlators, we see that the disconnected diagrams contributing at tree level give the result 
\begin{equation}\label{relationmixedsingle}
 \langle {\cal O}_q {\cal O}_q {\cal O}_p^2{\cal O}_p^2\rangle = 
4 
 \langle {\cal O}_q {\cal O}_q {\cal O}_p {\cal O}_p\rangle \langle {\cal O}_p {\cal O}_p \rangle + \ldots
\end{equation}
Normalizing we find
\begin{equation}
\frac{ \langle {\cal O}_q {\cal O}_q {\cal O}_p^2{\cal O}_p^2 \rangle}{ 
\langle {\cal O}_q {\cal O}_q\rangle \langle {\cal O}^2_p {\cal O}^2_p\rangle }= 
4 \frac{ \langle {\cal O}_p {\cal O}_p \rangle\langle {\cal O}_p {\cal O}_p \rangle }{ \langle {\cal O}^2_p {\cal O}^2_p\rangle}
\frac{  \langle {\cal O}_q {\cal O}_q {\cal O}_p {\cal O}_p\rangle }{ \langle {\cal O}_q {\cal O}_q\rangle \langle {\cal O}_p {\cal O}_p\rangle    } + \ldots
\end{equation}
By using that ${\cal P}_{qq[p^2][p^2]}= g_{34}^p {\cal P}_{qqpp}$, and that
\begin{equation}
    \langle {\cal O}_p^2 {\cal O}^2_p\rangle
= 2 \langle {\cal O}_p {\cal O}_p\rangle^2 \qquad {\rm at\ leading\ order}
\end{equation} 
we immediately read off the dynamical functions
\begin{align}
{\cal H}_{[2^2][2^2]44}\Big|_{\frac{1}{N^2}} =  2 {\cal H}_{2244}\Big|_{\frac{1}{N^2}}
\qquad;\qquad
{\cal H}_{22[3^2][3^2]}\Big|_{\frac{1}{N^2}}  = 2  {\cal H}_{2233}\Big|_{\frac{1}{N^2}}
\end{align}
For the case of the double particle operator ${\cal O}_4{\cal O}_2$ we will assume a linear combination of the form
\begin{align}
{\cal H}_{ 22[42][42] }= c_1 {\cal H}_{2222}+ c_2 {\cal H}_{2244}
\end{align}
and fix the coefficients $c_1,c_2$ from the OPE analysis. We will find that $c_1=c_2=1$.\\

It will be useful to recall that
\begin{equation}
{\cal H}_{22qq}\Big|_{\frac{1}{N^2}}= 
  \frac{-2q}{(q-2)!}(x_1 x_2)^{q}\overline{D}_{q,q+2,2,2}(x_1,x_2)\ .
\end{equation}
This gives the explicit expression for ${\cal H}_{22[2^2][2^2]},{\cal H}_{22[3^2][3^2]}$ and ${\cal H}_{22[42][42]}$. 
Then, cancellation of stringy states in $[000]$ becomes an almost trivial computation, since it is 
the same as for the single particle correlators times a factor of $2$ that comes from \eqref{relationmixedsingle}. 
For the case of ${\cal H}_{22[42][42]}$ we finds one relation between $c_1$ and $c_2$, which is $2c_1+c_2=3$.\\

As pointed out in the main text, when we analyze the crossed orientation, 
e.g.~${\cal H}_{[2^2]4[2^2]4}$, etc\ldots, the mixed single- and double-particle dynamical functions 
transform differently from their single-particle primitives, and this is crucial. 
For convenience, let us record here the expressions of
\begin{align}
\label{dyna3x3_1}
{\cal H}_{[2^2]4[2^2]4}&= 
-\frac{8}{(y_1y_2)^2} (x_1 x_2)^4 \overline{D}_{2,6,2,4} 
\rule{1cm}{0pt}\qquad;\qquad \rule{.9cm}{0pt}
{\cal P}_{4,4,4,4}=g_{12}^4 g_{34}^4\\[.2cm]
{\cal H}_{2[3^2]2[3^2]}&= -12 (x_1 x_2)^2 \overline{D}_{2,5,2,3} 
\rule{1.9cm}{0pt}\qquad;\qquad \rule{.9cm}{0pt}{\cal P}_{2,6,2,6}=g_{12}^2 g_{24}^4 g_{34}^2\\
{\cal H}_{2[42]2[42]}&= -4c_1 (x_1x_2)^2 \overline{D}_{2,4,2,2} - 4 c_2 (x_1x_2)^2 \overline{D}_{2,6,2,4} 
\label{dyna3x3_3}
\end{align}
Note that the dynamical functions here use the same $\overline{D}$ function as for the 
single particles correlators, but other details are different because crossing transformations 
depend on the external charges.
At this point the study of multiplet recombination in the $[040]$ rep is either a non trivial 
consistency check or it provides bootstrap constraints (with infinite spin support) through 
the cancellation of stringy states in \eqref{A=S+K-K}. \\

 There are two double particle  semishort operators in the $[040]$ rep, which 
 schematically have the form ${\cal O}_4\partial^l {\cal O}_2, {\cal O}_3\partial^l{\cal O}_3$. 
There are also multi-particle operators in this rep, but at leading order the double 
particles are the only ones that enter the OPE of our correlators. 
Then we can assemble the Gram matrix
\begin{equation}
\left( \begin{array}{cc|c} 
\langle {\cal O}_2{\cal O}_4  {\cal O}_2{\cal O}_4 \rangle & \langle {\cal O}_2{\cal O}_4  {\cal O}_3{\cal O}_3 \rangle &  \langle {\cal O}_2{\cal O}_4  O_{p_3} O_{p_4} \rangle \\[1.5ex]
 \langle {\cal O}_2{\cal O}_4  {\cal O}_3{\cal O}_3 \rangle &  \langle {\cal O}_3{\cal O}_3  {\cal O}_3{\cal O}_3 \rangle &  \langle {\cal O}_3{\cal O}_3  O_{p_3} O_{p_4} \rangle  \\[1.5ex]
 \hline
 %& &  \\
\rule{0pt}{0.5cm} \langle   O_{p_1} O_{p_2} {\cal O}_2{\cal O}_4 \rangle &  \langle O_{p_1} O_{p_2}  {\cal O}_3{\cal O}_3   \rangle  & \langle O_{p_1} O_{p_2} O_{p_3} O_{p_4} \rangle 
 \end{array}\right)
\end{equation}
and find the twist six semishort contribution to
$\langle O_{p_1} O_{p_2} O_{p_3} O_{p_4} \rangle$ by requiring that 
the determinant of this Gram matrix vanishes. 
The $2\times 2$ minor is $O(1)$ on the diagonal, thus the diagonals are 
leading in the computation we are interested in. Let's see what happens for the other entries. 

It is simple to see that for both 
$O_{p}O_{q}={\cal O}_2^2{\cal O}_4$ and  $O_{p}O_{q}={\cal O}_2[{\cal O}_2{\cal O}_4]$ 
the leading contribution in the Gram matrix comes from $\langle O_{p}O_{q} {\cal O}_2 {\cal O}_4\rangle$ through the diagrams
\begin{equation}
\begin{array}{ccccc}
\!\!\!\!\!\begin{tikzpicture}[scale=1.5]  
\def\shift{.35}

\def\latoxuno{-.35}
\def\latoxdue{-.38+0.8}
\def\latoyuno{.3}
\def\latoydue{-.25}

\draw[fill=red!60] 
(\latoxuno+.05, \latoydue) -- (\latoxdue, \latoyuno-.05) --  
(\latoxdue-.05, \latoyuno) -- (\latoxuno, \latoydue+.05) -- cycle;

\draw (\latoxuno-.1, \latoyuno) --  (\latoxuno-.1, \latoydue+.1);
\draw (\latoxuno-.1, \latoyuno) --  (\latoxdue, \latoyuno);

\draw (\latoxdue-.05, \latoydue) --  (\latoxuno, \latoyuno+.2-.05);
\draw (\latoxdue, \latoydue+.05) --  (\latoxuno+.05, \latoyuno+.2);

\draw[fill=white!60,draw=white] (\latoxuno-.2, \latoyuno-.1) rectangle (\latoxuno, \latoyuno+.1);
\draw[fill=white!60,draw=white] (\latoxdue-.1, \latoydue-.05) rectangle (\latoxdue+.15, \latoydue+.1);

    \draw[fill=white!60,draw=white] (\latoxdue+.02, \latoyuno+.02) circle (2.5pt);
    \draw[fill=white!60,draw=white] (\latoxuno-.02, \latoydue+.02) circle (2.5pt);

\draw(\latoxuno-.1, \latoyuno+.2+.08) node[scale=.8] {${\cal O}_2$};
   \draw(\latoxuno-.2, \latoyuno+.08) node[scale=.8] {${\cal O}_2$};
   \draw(\latoxuno-.04, \latoydue-.08) node[scale=.8] {${\cal O}_4$};
   \draw(\latoxdue+.14, \latoyuno+.03) node[scale=.8] {${\cal O}_4$};
   \draw(\latoxdue+.1, \latoydue-.05) node[scale=.8] {${\cal O}_2$};

 \draw(\latoxdue+1,\latoydue+.2) node[scale=1] {$;$};
\end{tikzpicture} 
&
\qquad \qquad 
&
\!\!\!\!\!\begin{tikzpicture}[scale=1.5]  
\def\shift{.35}

\def\latoxuno{-.35}
\def\latoxdue{-.38+0.8}
\def\latoyuno{.3}
\def\latoydue{-.25}

\draw (\latoxuno-.5,0.05) node[scale=1] {$\displaystyle 8$};

\draw[fill=red!60] 
(\latoxuno, \latoyuno+.035) -- (\latoxdue-.035, \latoyuno+.035) --  
(\latoxdue-.035, \latoyuno-.035) -- (\latoxuno, \latoyuno-.035) -- cycle;

\draw (\latoxuno-.1, \latoyuno) --  (\latoxuno-.1, \latoydue+.1);
\draw (\latoxuno, \latoydue) --  (\latoxdue, \latoyuno);

\draw (\latoxdue-.05, \latoydue) --  (\latoxuno, \latoyuno+.2-.05);
\draw (\latoxdue, \latoydue+.05) --  (\latoxuno+.05, \latoyuno+.2);

\draw[fill=white!60,draw=white] (\latoxuno-.2, \latoyuno-.1) rectangle (\latoxuno, \latoyuno+.1);
\draw[fill=white!60,draw=white] (\latoxdue-.1, \latoydue-.05) rectangle (\latoxdue+.15, \latoydue+.1);

    \draw[fill=white!60,draw=white] (\latoxdue+.02, \latoyuno+.02) circle (2.5pt);
    \draw[fill=white!60,draw=white] (\latoxuno-.02, \latoydue+.02) circle (2.5pt);

\draw(\latoxuno-.1, \latoyuno+.2+.08) node[scale=.8] {${\cal O}_2$};
   \draw(\latoxuno-.2, \latoyuno+.08) node[scale=.8] {${\cal O}_4$};
   \draw(\latoxuno-.04, \latoydue-.08) node[scale=.8] {${\cal O}_2$};
   \draw(\latoxdue+.14, \latoyuno+.03) node[scale=.8] {${\cal O}_4$};
   \draw(\latoxdue+.1, \latoydue-.05) node[scale=.8] {${\cal O}_2$};

\end{tikzpicture}
&
&
\!\!\!\!\!\begin{tikzpicture}[scale=1.5]  
\def\shift{.35}

\def\latoxuno{-.35}
\def\latoxdue{-.38+0.8}
\def\latoyuno{.3}
\def\latoydue{-.25}

\draw (\latoxuno-.6,0.05) node[scale=1] {$\displaystyle+\ 4$};

\draw[fill=red!60] 
(\latoxuno, \latoyuno+.035) -- (\latoxdue-.035, \latoyuno+.035) --  
(\latoxdue-.035, \latoyuno-.035) -- (\latoxuno, \latoyuno-.035) -- cycle;

\draw (\latoxuno-.1, \latoyuno) --  (\latoxuno-.1, \latoydue+.1);
\draw (\latoxuno, \latoydue) --  (\latoxdue, \latoydue);

\draw (\latoxdue, \latoydue) --  (\latoxuno, \latoyuno+.2-.05);

\draw[fill=white!60,draw=white] (\latoxuno-.2, \latoyuno-.1) rectangle (\latoxuno, \latoyuno+.1);
\draw[fill=white!60,draw=white] (\latoxdue-.1, \latoydue-.05) rectangle (\latoxdue+.15, \latoydue+.1);

    \draw[fill=white!60,draw=white] (\latoxdue+.02, \latoyuno+.02) circle (2.5pt);
    \draw[fill=white!60,draw=white] (\latoxuno-.02, \latoydue+.02) circle (2.5pt);

\draw(\latoxuno-.1, \latoyuno+.2+.08) node[scale=.8] {${\cal O}_2$};
   \draw(\latoxuno-.2, \latoyuno+.08) node[scale=.8] {${\cal O}_4$};
   \draw(\latoxuno-.04, \latoydue-.08) node[scale=.8] {${\cal O}_2$};
   \draw(\latoxdue+.14, \latoyuno+.03) node[scale=.8] {${\cal O}_4$};
   \draw(\latoxdue+.1, \latoydue-.05) node[scale=.8] {${\cal O}_2$};

\end{tikzpicture}
\end{array}
\end{equation}
while in the case of $O_{p}O_{q}={\cal O}_2{\cal O}_3^2$ the only contribution comes from $\langle O_{p}O_{q}{\cal O}_3{\cal O}_3\rangle$ through the diagrams
\begin{equation}
\begin{array}{cc}
\!\!\!\!\!\begin{tikzpicture}[scale=1.5]  
\def\shift{.35}

\def\latoxuno{-.35}
\def\latoxdue{-.38+0.8}
\def\latoyuno{.3}
\def\latoydue{-.25}

\draw[fill=red!60] 
(\latoxuno, \latoyuno+.035) -- (\latoxdue-.035, \latoyuno+.035) --  
(\latoxdue-.035, \latoyuno-.035) -- (\latoxuno, \latoyuno-.035) -- cycle;

\draw (\latoxuno-.1, \latoyuno) --  (\latoxuno-.1, \latoydue+.1);
\draw (\latoxuno, \latoydue) --  (\latoxdue, \latoyuno);

\draw (\latoxdue-.05, \latoydue) --  (\latoxuno, \latoyuno+.2-.05);
\draw (\latoxdue+.01, \latoydue) --  (\latoxuno+.01, \latoyuno+.2);
\draw (\latoxdue+.03, \latoydue+.05) --  (\latoxuno+.03, \latoyuno+.2+.05);

\draw[fill=white!60,draw=white] (\latoxuno-.2, \latoyuno-.1) rectangle (\latoxuno, \latoyuno+.1);
\draw[fill=white!60,draw=white] (\latoxdue-.1, \latoydue-.05) rectangle (\latoxdue+.15, \latoydue+.1);

    \draw[fill=white!60,draw=white] (\latoxdue+.02, \latoyuno+.02) circle (2.5pt);
    \draw[fill=white!60,draw=white] (\latoxuno-.02, \latoydue+.02) circle (2.5pt);

\draw(\latoxuno-.1, \latoyuno+.2+.08) node[scale=.8] {${\cal O}_3$};
   \draw(\latoxuno-.2, \latoyuno+.08) node[scale=.8] {${\cal O}_3$};
   \draw(\latoxuno-.04, \latoydue-.08) node[scale=.8] {${\cal O}_2$};
   \draw(\latoxdue+.14, \latoyuno+.03) node[scale=.8] {${\cal O}_3$};
   \draw(\latoxdue+.1, \latoydue-.05) node[scale=.8] {${\cal O}_3$};

\end{tikzpicture} 
&
\qquad
\!\!\!\!\!\begin{tikzpicture}[scale=1.5]  
\def\shift{.35}

\def\latoxuno{-.35}
\def\latoxdue{-.38+0.8}
\def\latoyuno{.3}
\def\latoydue{-.25}

\draw (\latoxuno-.9,0.05) node[scale=1] {$\displaystyle+$};

\draw[fill=red!60] 
(\latoxuno+.015, \latoyuno+.035) -- (\latoxuno-.015, \latoyuno-.035) --  
(\latoxdue-.015, \latoydue-.035) -- (\latoxdue+.015, \latoydue+.035) -- cycle;

\draw (\latoxuno-.1, \latoyuno) --  (\latoxuno-.1, \latoydue+.1);
\draw (\latoxuno, \latoydue) --  (\latoxdue, \latoydue);

\draw (\latoxdue-.04, \latoyuno) --  (\latoxuno, \latoyuno+.2-.04);
\draw (\latoxdue, \latoyuno+.04) --  (\latoxuno+.02, \latoyuno+.2);
\draw (\latoxdue+.04, \latoyuno+.08) --  (\latoxuno+.02, \latoyuno+.2+.05);

 \draw[fill=white!60,draw=white] (\latoxdue, \latoydue) circle (2.5pt);
    \draw[fill=white!60,draw=white] (\latoxuno-.1, \latoyuno+.05) circle (2.5pt);

    \draw[fill=white!60,draw=white] (\latoxdue+.02, \latoyuno+.02) circle (2.5pt);
    \draw[fill=white!60,draw=white] (\latoxuno-.02, \latoydue+.02) circle (2.5pt);

\draw(\latoxuno-.1, \latoyuno+.2+.08) node[scale=.8] {${\cal O}_3$};
   \draw(\latoxuno-.2, \latoyuno+.08) node[scale=.8] {${\cal O}_3$};
   \draw(\latoxuno-.04, \latoydue-.08) node[scale=.8] {${\cal O}_2$};
   \draw(\latoxdue+.14, \latoyuno+.03) node[scale=.8] {${\cal O}_3$};
   \draw(\latoxdue+.1, \latoydue-.05) node[scale=.8] {${\cal O}_3$};

\end{tikzpicture}
\end{array}
\end{equation}
At this point we ready for the computation. 
The normalization of the double-particle operators are
\begin{equation}
{\cal N}_{[2^2]}= 2\, {\cal N}_2^{\,2} \left(1+\frac{2}{a}\right)\qquad;\qquad{\cal N}_{[42]}= {\cal N}_2{\cal N}_4 \left(1+\frac{8}{a}\right) \qquad;\qquad {\cal N}_{[3^2]}=\,2\,{\cal N}^{\,2}_{3}\, \left( 1+\frac{9}{a}\right)\,.
\end{equation}
and the free theories are 
\begin{align}
\label{suppl2d4240}
\frac{ \langle{\cal O}_{2}^2{\cal O}_{4} {\cal O}_2 {\cal O}_4 \rangle}{{\cal N}_{[2^2]}{\cal N}_4} &= g_{12}^3 g_{14}g_{24}g_{34}^2 \Bigg[ \frac{4U^2\sigma^2}{a} + \frac{24}{a(a+2)}\Bigg[ U\sigma +\frac{U\tau}{V} +\frac{U^2\tau\sigma}{V} \Bigg]\Bigg] \\[.2cm]
\label{suppl2d4241}
\frac{ \langle  [{\cal O}_{4}{\cal O}_2] {\cal O}_2 {\cal O}_2 {\cal O}_4 \rangle}{ {\cal N}_{[42]} } &=g_{12}^2g_{13}g_{14}^3g_{34}\Bigg[ 8 U\sigma + \frac{4U\tau}{V}+8\Bigg]\\[.2cm]
\label{suppl2d4242}
\frac{ \langle {\cal O}_{3}^2  {\cal O}_{2}  {\cal O}_3 {\cal O}_3   \rangle}{ {\cal N}_{[3^2]} }&= g_{12}^2g_{13}^2g_{14}^2 g_{34} \Bigg[ 6U\sigma+\frac{6U\tau}{V} + \frac{ 36(a-7)}{(a-3)(a+9) } \Bigg]
\end{align}
The predictions for the twist six semishort contribution are all non trivial functions of the spin:
\begin{align}
{\cal S}^{2^242^24}_{[040]}&=4\frac{ \big(4 M^{4,4, 2,4}_{0,6,[2+l]}\big)^2}{  M^{2,4,2,4}_{0,6,[2+l]}  }\\ %\quad;\quad 
{\cal S}^{2[42]2[42]}_{[040]}&=\frac{1}{2}\frac{\big( 8 M^{6,2,2,4}_{0,6,[2+l]} + 4 M^{6,2,2,4}_{1,6,[2+l]} \big)^2 }{ M^{2,4,2,4}_{0,6,[2+l]} } \\
%\quad;\quad
{\cal S}^{23^223^2}_{[040]}&=\frac{ \big(6 M^{6,2, 3,3}_{0,6,[2+l]}+6 M^{6,2, 3,3}_{1,6,[2+l]} \big)^2}{  M^{3,3,3,3}_{0,6,[2+l]} +M^{3,3,3,3}_{3,6,[2+l]}  } 
\end{align}
The prefactor outside parenthesis come from compensating the normalizations that 
we used to write the various correlators in (\ref{suppl2d4240}-\ref{suppl2d4242}). For example 
in  ${\cal S}^{2^242^24}_{[040]}$, the factor $4$ is just $4=({\cal N}_{[2^2]} {\cal N}_{4})^2/ ({\cal N}_2 {\cal N}_4)/({\cal N}_{[2^2]}{\cal N}_4)$, etc\ldots. \\

To conclude, we will compute the stringy contributions 
in each case, by solving \eqref{A=S+K-K}. The free theories are
\begin{align}
\!\!\!\!\!\!\!\!\!\!\langle {\cal O}_2^2  {\cal O}_4 {\cal O}_2^2 {\cal O}_4 \rangle=\ & {\cal N}_{[2^2]}{\cal N}_{4} g_{12}^4g_{34}^4\Bigg[ U^4\sigma^4 +\frac{16}{a}\frac{U^4\sigma^3\tau}{V}+ \frac{16}{a} U^3\sigma^3  +\frac{48}{(a+2)}\frac{U^3\sigma^2\tau}{V} + \\
& +\frac{48}{a(a+2)}\Bigg[ \frac{ U^4\sigma^2\tau^2}{V^2}+\frac{6U^3\sigma^2\tau}{V} +\frac{4U^3\sigma\tau^2}{V^2} + \frac{48}{a(a+2)}  U^2\sigma^2 + \frac{4U^2\sigma\tau}{V}+\frac{U^2\tau^2}{V^2} \Bigg] \Bigg]\notag\\[.2cm]
\!\!\!\!\!\!\!\!\!\!\langle {\cal O}_2  [{\cal O}_4{\cal O}_2] {\cal O}_2  [{\cal O}_4{\cal O}_2]\rangle=\ & {\cal N}_{2}{\cal N}_{[42]}g_{12}^2 g_{24}^4 g_{34}^2\Bigg[ U^2\sigma^2 + \left( 1+\frac{8}{a}\right)\Bigg[ 1 + \frac{U^2\tau^2}{V^2}\Bigg] + \frac{12}{a} U\sigma+ \frac{60}{a} \frac{U\tau}{V} + \frac{12}{a} \frac{U^2\sigma\tau}{V}\Bigg] \\[.2cm]
\!\!\!\!\!\!\!\!\!\!\langle {\cal O}_2  {\cal O}_3^2  {\cal O}_2{\cal O}_3^2\rangle=\ & {\cal N}_{2}{\cal N}_{[3^2]}g_{12}^2 g_{24}^4 g_{34}^2\Bigg[ U^2\sigma^2 + \frac{36}{a}\frac{(a-7)}{(a-3)(a+9)}\Bigg[ 1+\frac{U^2\tau^2}{V^2}\Bigg]+ \frac{12}{a}U\sigma + \frac{60}{a}\frac{U\tau}{V} +\frac{12}{a}\frac{U^2\sigma\tau}{V} \Bigg]
\end{align}
Again we only need to use the $M$aster formula in each case, to solve for ${\cal K}_{[040]}$. For example,
\begin{equation}
{\cal K}^{2[42]2[42]}_{[040]}\Big|_{\frac{1}{a}}= 12 M^{2,6,2,6}_{0,6,[2+l]}+ 60 M^{2,6,2,6}_{1,6,[2+l]}- {\cal S}^{2[42]2[42]}_{[040]}
\end{equation}
At this point it is rather immediate to show that these predictions cancels again 
the twist six contribution coming from the corresponding dynamical functions 
in \eqref{dyna3x3_1}-\eqref{dyna3x3_3}. In the case of ${\cal H}_{2[42]2[42]}$ the only 
solution is $c_1=c_2=1$, which agrees with the constraint derived from $[000]$, namely $2c_1+c_2=3$.

%=====================================================================
\subsection{CFT data for $\langle {\cal O}_2^2 {\cal O}_q {\cal O}_2^2 {\cal O}_q\rangle$ at tree level.}

So far we have analyzed aspects of the tree level correlators ${\cal H}_{[2^2][2^2]22}$, ${\cal H}_{[2^2][2^2]33}$ 
and ${\cal H}_{[2^2][2^2]44}$. We will now discuss the general case of ${\cal H}_{[2^2][2^2]qq}$, for which we claim that 
${\cal H}_{[2^2][2^2]qq}= 2 {\cal H}_{22qq}$ is the dynamical function. 
While the block expansion in this channel satisfies the constraints from the absence of 
stringy states in an obvious way, i.e.~because the single particle correlator $ {\cal H}_{22qq}$ does, 
the analysis in the crossed channel ${\cal H}_{[2^2]q[2^2]q}$ is quite non trivial.\\  

Let us start from quoting some explicit results,
\begin{align}
 M^{2^2\!q2^2\!q}_{0,q+2,[2+l]}&= \frac{(-1)^l}{2(q-2)!}(l+3)(l+4)(l+q+1)(l+q+2) \frac{(l+q)!^2}{(2l+q+4)!} \\
 M^{2^2\!q2^2\!q}_{1,q+2,[2+l]}&=  \frac{(l+4)!(l+q)!}{(2l+q+4)!}+\frac{(-1)^l}{(q-3)!}\left[ - \frac{(q-3)(q+1)}{4} +\left(l+\frac{q+5}{2}\right)^{\!2}\right] \frac{(l+q)!^2}{(2l+q+4)!} \\[.2cm]
{\cal S}^{2^2\!q2^2\!q}_{[0q0]}\Big|_{\frac{1}{N^2}}&= \frac{4 q^2(q-1)}{(q-2)!} (-1)^l (l+3)(l+q+2) \frac{(l+q)!^2}{(2l+q+4)!}
\end{align}
Putting these results together we solve \eqref{A=S+K-K} and find the stringy contribution,
\begin{align}
&
{\cal K}^{2^2\!q2^2\!q}_{[0q0]}\Big|_{\frac{1}{N^2}}=4q \left[ M^{2^2\!q2^2\!q}_{0,q+2,[2+l]} + (q-1)M^{2^2\!q2^2\!q}_{1,q+2,[2+l]}\right] - {\cal S}^{2^2\!q2^2\!q}_{[0q0]}\Big|_{\frac{1}{N^2}}\\
& =\frac{ 
 (l+\frac{q+5}{2})^4 - \frac{1}{2}( (q+2)^2-7) (l+\frac{q+5}{2})^2 + \frac{(q-1)}{16} ( (q+3)^3-12(q+7) )} {(-1)^l(q-2)!} \frac{2q(l+q)!^2}{(2l+q+4)!}+  4q(q-1)\frac{(l+4)!(l+q)!}{(2l+q+4)!} \notag
\end{align}
In a similar way one computes and finds the relations
\begin{align}
{\cal K}^{2^2\!q2^2\!q}_{[1,q-2,1]}\Big|_{\frac{1}{N^2}}=& 4q \left[ M^{2^2\!q2^2\!q}_{0,q+2,[2+l,1]} + (q-1)M^{2^2\!q2^2\!q}_{1,q+2,[2+l,1]}\right]-{\cal S}^{2^2\!q2^2\!q}_{[1,q-2,1]}\Big|_{\frac{1}{N^2}}=
\frac{2(q-2)}{-q}{\cal K}^{2^2\!q2^2\!q}_{[0,q,0]}\Big|_{\frac{1}{N^2}} \\[.2cm]
{\cal K}^{2^2\!q2^2\!q}_{[2,q-4,2]}\Big|_{\frac{1}{N^2}}=&
4q \left[ M^{2^2\!q2^2\!q}_{0,q+2,[2+l,1,1]} + (q-1)M^{2^2\!q2^2\!q}_{1,q+2,[2+l,1,1]}\right]=
\frac{q-3}{q-1}{\cal K}^{2^2\!q2^2\!q}_{[0,q,0]}\Big|_{\frac{1}{N^2}}
\end{align}
Note that the $M$ coefficients and the ${\cal S}$ coefficients are all different, 
and therefore the above relations with ${\cal K}^{2^2\!q2^2\!q}_{[0,q,0]}$ are very non trivial!\\

In order to check in each case that the stringy contributions are cancelled, 
we follow \cite{Aprile:2017xsp}. To this end it will be convenient to rewrite the dynamical functions as
\begin{align}
{\cal H}_{22^222^2}&= %{\cal P}_{22^222^2}\, {\cal I}\Bigg[ 
-\frac{8}{(y_1y_2)^0} (x_1 x_2)^2 \overline{D}_{2,4,2,2} %\Bigg] 
\rule{.9cm}{0pt}
\qquad;\qquad 
{\cal P}_{2,4,2,4}=g_{12}^2 g_{24}^{2} g_{34}^2
\\
{\cal H}_{32^232^2}&= %{\cal P}_{32^232^2}\, {\cal I} \Bigg[ 
-\frac{12}{(y_1y_2)^1} (x_1 x_2)^3 \overline{D}_{2,5,2,3} %\Bigg] 
\rule{.9cm}{0pt}
\qquad;\qquad 
{\cal P}_{3,4,3,4}=g_{12}^3 g_{24}^{} g_{34}^3
\\
{\cal H}_{2^2q2^2q}&= %{\cal P}_{2^2q2^2q} \,{\cal I} \Bigg[ 
-\frac{4q/(q-2)!}{(y_1y_2)^2} (x_1 x_2)^4 \overline{D}_{2,q+2,2,q} 
%\Bigg] 
\qquad;\qquad 
{\cal P}_{4,q,4,q}=g_{12}^4 g_{24}^{p-4} g_{34}^4
\end{align}
The blocks for the long sector are given in \cite[(74)]{Aprile:2017xsp}.\\

%

%=========================================================================================
\section{Determinant formulae for all free theories}
%=========================================================================================

In the main text we have obtained a master formula for the block decomposition of the most general graph $G$  
that contributes to a four-point function with external scalar operators. 
In this case the graph $G$ is built out of products of scalar propagators as
 displayed in~\eqref{CFT_equa}. In the previous sections we discussed applications to ${\cal N}=4$ SYM 
 in four dimensions, but we would like to emphasise that our formula can be used in many different context. 
 For example, the $\alpha$-space identity of \cite{Hogervorst:2017sfd}, recently 
 used in \cite{Poland:2024rxv} in the study of correlators with insertions of maximally heavy operators, reads, 
\begin{equation}\label{alphaid}
x^{p}(1-x)^{-k}=\sum_{j=0}^{\infty} M_{k,\gamma=2p,\ulmb} \Bigg[ x^{\Delta}\,_2F_1\Big[\,^{\Delta,\Delta}_{\ 2\Delta};x \Big]\Bigg]
\qquad;\qquad ,\ \Delta=p+j\ ,\ \ulmb=[j]
\end{equation}
For $k\leq p$ we immediately recognize this identity as a one-dimensional block 
expansion, thus from \eqref{superCauchy2} we get
\begin{equation}\label{superCauchy2app}
\frac{1}{\!\!\prod_{\l=1}^{d} 
		(1-Y_{\l} x)}=  	
	\sum_{\ulmb } 
	 J^{(p)}_\ulmb(Y) \,
		{F}^{(1,0)}_{\gamma=2p,\,\ulmb}(x)\,\qquad;\qquad Y=[1^k,0^{p-k}]
\end{equation}
Then from our master formula one can check that the combinatorics of our \eqref{Acoeffs} reproduces 
\begin{equation}\label{poland_formula}
M_{k,\gamma,[j]} = \frac{(\frac{\gamma}{2})_j(\frac{\gamma}{2})_j }{j!(\gamma+j-1)_j}\,_3F_2\Big[\,^{-j,-k+\frac{\gamma}{2},-1+j+\gamma}_{\rule{1cm}{0pt}\frac{\gamma}{2},\frac{\gamma}{2}};1\Big]
\end{equation}
which is precisely the result quoted in \cite[(4.14)]{Poland:2024rxv}. The blocks in \eqref{alphaid} 
are simple enough that one can perform the block decomposition and find directly the $M_{k,2p,[j]}$. 
Nevertheless, our main point here is that there is no need to perform the block decomposition explicitly, 
rather, the master formula gives the block coefficients directly, allowing one to focus on the physical 
properties of these numbers, e.g.~study the dependence on parameters. Like this there are many 
other examples where our formula is directly applicable. One last thing to note about the RHS 
of \eqref{poland_formula} is that it can be analytically continued for any $\gamma,k$, thus outside the 
combinatorial definition of the Jacobi polynomials. In the following we will see what is the 
counterpart of this continuation in our formalism.\\ 

Besides the case of theories where the elementary fields are scalar fields, we would like 
to include theories where the elementary fields are fermions or vector fields, and study 
correlation functions of scalar operators built out these elementary fields. For example  
in a 4d CFT,  we could consider the dimension three (scalar) operator $\cO_3 =\psi  \psi'$, 
built out of fermion fields, and compute the correlator $\langle   \cO_3 \bar \cO_3 \cO_3 \bar \cO_3 \rangle$. 
The fermion propagator is $\langle \psi_\alpha \bar \psi_{\dot \alpha}\rangle \sim (x_{12})_{\alpha \dot \alpha}/x_{12}^4$, 
and we would find the expression
\begin{align}\label{fermcor}
\langle   \cO_3 \bar \cO_3 \cO_3 \bar \cO_3 \rangle =a_1 \Big(g_{12}^3g_{34}^3+g_{14}^3g_{23}^3\Big) +a_2 \Big(g_{12}g_{34}g_{14}^2g_{23}^2+g_{12}^2g_{34}^2g_{14}g_{23} - \frac{g_{12}^2g_{34}^2g_{14}^2g_{23}^2}{g_{13}g_{24}}\Big)
\end{align}
where $a_i$ are numbers obtained from Wick's theorem. Note, the second term arises from 
the cyclic Wick contraction yielding $\tr(x_{12}x_{23}x_{34}x_{41}) \sim x_{12}^2x_{34}^2+x_{14}^2x_{23}^2-x_{13}^2x_{24}^2$. 
This last term has inverse powers of $g_{ij}$, and thus falls outside the cases 
for which we derived the master formula in \eqref{Acoeffs}.
The case of scalar operator $\frac{1}{4} F^2$, built from the field strength of 
a free vector field has similar features. 
See e.g. \cite[(6.27) and (6.34)]{Dolan:2000ut}.\\

In order to generalize our master formula we will consider free theory four point functions 
that admit an expansion  as in \eqref{intro_4pt}, namely
\begin{equation}\label{generalfree}
\langle \cO_1 \cO_2 \cO_3 \cO_4 \rangle = 
{\cal P}\times \sum_{\gamma,k} a_{G_{\gamma,k}}\,U^{\frac{\gamma}{2}} V^{-k}
\qquad;\qquad U=\frac{ g_{13} g_{24} }{ g_{12} g_{34} }\quad;\quad V= 
\frac{ g_{13} g_{24} }{g_{14} g_{23} }
\end{equation}
but where the notion of a graph $G$ is generalized so that in the expression
$G={\cal P}\times U^{\frac{\gamma}{2}}V^{-k}$  we allow $k$ to take values also 
in the range $k<0$ and $k>\beta=\frac{1}{2}(\gamma-\gamma_{\min})$. 
Note that, as in the example above, \eqref{fermcor}, the values $k<0$ and $k>\beta$ 
simply arise from the fact that fundamental propagators of fermions and vectors have Lorentz structure. 
The sums over $\gamma$ and $k$ are still finite! \\

Although not falling within the master formula \eqref{Acoeffs}, we will now show that we can compute 
the block coefficient of any generalized graph $G$ from the more general Cauchy identity \cite[(2.30),(8.7)]{Aprile:2021pwd}
\begin{equation}\label{superCauchy3}
\frac{\,\prod_{\l,j}^{\mathsf{M},n} (1-Y_{\l} y_j)\prod_{\l,i}^{\mathsf{N},m} 
		(1-X_{\l} x_i)}{\!\!\prod_{\l,k}^{\mathsf{M},m} 
		(1-Y_{\l} x_i)  \prod_{\l,j}^{\mathsf{N},n} 
		(1-X_{\l}y_j)^{}} =  	
	\sum_{\ulmb } 
	 J^{(\mathsf{M},\mathsf{N})}_\ulmb(Y,X) \,
		{F}^{(m,n)}_{\gamma,\,\ulmb}(x,y)\qquad;\qquad \tfrac{1}{2}\gamma=\mathsf{M}-\mathsf{N} + \tfrac{1}{2}\gamma_{min}
\end{equation}
where $J_{\ulmb}^{(\mathsf{M},\mathsf{N})}(Y,X)$ are the superJacobi polynomials introduced 
by Veselov and Sergeev in \cite{Veselov_3,Veselov_4}. The identity \eqref{superCauchy3} 
reduces to \eqref{superCauchy2} for $\mathsf{N}=0$ and $\mathsf{M}$ set equal to the value 
of $d$ in \eqref{superCauchy2}. In particular, the super Jacobi polynomials with $\mathsf{N}=0$ become 
the Jacobi polynomials quoted in \eqref{superCauchy2}. \\

The logic of our proof goes through as follows. For each monomial in \eqref{generalfree} 
we want to match first the LHS of \eqref{superCauchy3}.
To achieve this note that the power of $U$ determines the value of $\mathsf{M}-\mathsf{N}$ 
through the relation $\mathsf{M}-\mathsf{N}=\frac{1}{2}\gamma-\frac{1}{2}\gamma_{min}=\beta$. Then,
in order to match $V^{-k}$ when $k<0$ we set $\mathsf{N}=-k$ and $X_{1,\ldots,\mathsf{N}}=1$. Since $\mathsf{M}=\beta-k$ we have to set $Y_{1,\ldots,\mathsf{M}}=0$.
In order to match $V^{-k}$ when $k>\beta$ we set $\mathsf{M}=k$ and $Y_{1,\ldots,\mathsf{M}}=1$. Since $\mathsf{N}=k-\beta$ we have to set $X_{1,\ldots,\mathsf{N}}=0$.
In both cases we reduce the computation of the block coefficients $M_{k,\gamma,\ulmb}$ to the 
evaluation of a super Jacobi polynomial at coincident values of $X,Y\in\{0,1\}$ with $X\neq Y$. Summarising
\begin{equation}\label{newGeneralizedG}
M_{k,\gamma,\ulmb}=\left\{\begin{array}{cc} 
J_{\ulmb}^{(\beta-k,-k)}(0^{\beta-k},1^{-k};p_{-},p_{+}) & k<0 \\[.3cm]
J_{\ulmb}^{(k,k-\beta)}(1^k,0^{k-\beta};p_{-},p_{+}) & k>\beta
\end{array}\right.
\end{equation}
in addition to the known scalar case 
\begin{equation}
M_{k,\gamma,\ulmb}=\ J_{\ulmb}^{(\beta,0)}([1^k,0^{\beta-k}];p_{-},p_{+})  \qquad 0\leq k\leq \beta 
\end{equation}
Compared to the scalar case, the new feature of \eqref{newGeneralizedG} is the wider 
range of values of $k$ (due to inverse propagators) that requires a {\em super} Jacobi polynomial 
where the $X$ variables are all set to  0   and the $Y$s are all set to 1, or vice versa, whereas 
in the scalar case (with no inverse propagators) it was enough to consider standard Jacobi 
polynomials where some of the $Y$ variables are set to 1 and some to 0. (Recall that the 
order is not important since the polynomial is symmetric.)

%===============================================================================
\subsection{Determinant formula for the superJacobi polynomials}

Following the steps of our proof, we should now find a non combinatorial formula for the super 
Jacobi polynomials. Remarkably, this can be done! Inspired by a determinantal representation of super 
Schur polynomials found in \cite{MVdJ}, we have found that the super Jacobi polynomials admit the representation,
\begin{align}
\!\!\!J^{(\mathsf{M},\mathsf{N})}_\ulmb=&\  
\frac{ \prod_{i,j}^{\mathsf{M},\mathsf{N}} (Y_i-X_j) }{{\rm VdM}(Y) {\rm VdM}(X) }
(-)^\sigma\!\det\!\left( \begin{array}{cc} \frac{1}{Y_i-X_j} & E_{\ulmb}(Y_i) 
\\[.2cm]
S_{\ulmb}(X_j) & 0\ \ \ \end{array} \right) \notag \\[.2cm]
\label{SJACOBI}
E_{\ulmb}=&\, \Big( J_{\lambda_l+\mathsf{M}-\mathsf{N}-l}(Y_i;\,+p^-,+p^+) \Big)_{\!\substack{1\leq i\leq \mathsf{M}\\[.1cm] \ \ 1\leq l\leq \kappa-1}} \\
S_{\ulmb}=&\, \Big( J_{\lambda'_l-\mathsf{M}+\mathsf{N}-l}(X_j;\,-p^-,-p^+) \Big)_{\!\!\!\substack{\ \ \, 1\leq l\leq \mathsf{N}-\mathsf{M}+\kappa-1\\[.1cm] \!\!\!\!\!\!\!\!\!\!\!\!\!\!\!\!\!\!\!1\leq j\leq \mathsf{N}}} \notag 
\end{align}
where 
\begin{align}
    \kappa=&\, \min\{i\,|\, \lambda_i+\mathsf{M}+1-i\leq \mathsf{N}\}\\[.2cm]
    \sigma=&\, \tfrac{(\kappa-1-\mathsf{M})(\kappa-2+\mathsf{M})}{2}+\sum_{j=1}^{\mathsf{N}-\mathsf{M}+\kappa-1}\lambda'_j
\end{align}
and the one-variable Jacobi polynomial is
\begin{equation}\label{univariateJ}
 J_{[\lambda]}(\textunderscore\textunderscore\,;a_1,a_2)= \frac{ (-)^\lambda  (a_1+1)_{\lambda}
 }{(a_1+a_2+1+\lambda)_\lambda } 
 \,_2F_1\Big[\,^{-\lambda,\,a_1+a_2+1+\lambda}_{\rule{.5cm}{0pt}\, a_1+1};\textunderscore\textunderscore\Big]\,.
 \end{equation}
Some comments on \eqref{SJACOBI} are in order. Note that $\lambda_{\mathsf{M}+1}\leq \mathsf{N}$ 
implies $1\leq \kappa\leq \mathsf{M}+1$. If $\kappa=\mathsf{M}+1$ then $\ulmb$ is typical and the 
determinant factorises as $\det E_{\ulmb}\times \det S_{\ulmb}$, but otherwise $\ulmb$ is atypical 
and the determinant does not factorise! When $\mathsf{M}=0$ or $\mathsf{N}=0$ there is no space 
for the matrix $1/(Y_i-X_j)$ and \eqref{SJACOBI} reduces to the known bosonic formula 
\cite[(7.2)]{Okounkov_Olshanski}, respectively, for $\ulmb'$ and $\ulmb$, that we used already 
in the main text, see \eqref{Jdet}. \\

For all cases in which $\mathsf{M},\mathsf{N}\neq 0$, the formula \eqref{SJACOBI} is new. 
Here below we review the original combinatorial definition super Jacobi polynomials \cite{Veselov_3,Veselov_4}, 
that allows a direct check of \eqref{SJACOBI} in examples. Some readers might wish 
to jump directly to next section where we discuss the corresponding formula for the 
generalized $M_{G,\gamma,\ulmb}$.\\ 

As in \eqref{SJACOBI} BC super Jacobi polynomials are labelled by a Young diagram $\ulmb$, 
and are specified by a number of variables, and two parameters, $a_1,a_2$ in the following.  
The general theory has an extra parameter $\theta$, which will be set to $\theta=1$ at the end. 
Super Jacobi polynomials can then be defined as a finite expansion over super 
Jack polynomials (that for $\theta=1$ are super Schur polynomials), 
\begin{equation}\label{jacobiinjack}
{J}_{\ulmb}({\bf s})= \sum_{\umu\subseteq\ulmb} (S_{\gamma})^{\umu}_{\ulmb}\, P_{\mu}({\bf s};\theta)\ .
\end{equation} 
It turns out that the coefficient $(S_{\gamma})^{\umu}_{\ulmb}$ is special. In the math literature it is 
known as ``binomial coefficient". Okounkov and Rains showed (in a more general context) 
that it can be written in terms of polynomials evaluated at partitions. 
From this statement one arrives at the formula \cite{Aprile:2021pwd},
\begin{equation}\label{Spaul}
(S_{\gamma})^{\umu}_{\ulmb}= (-)^{|\umu|-|\ulmb|} \frac{\Pi_{\ulmb}(\theta) }{ \Pi_{\umu}(\theta)}
C^0_{\umu\slash\ulmb}( \theta\alpha,\theta\beta;\theta)\frac{ {\tilde P}_{\ulmb}^{*}(\umu;\theta,h) }{  {\tilde P}_{\umu}^{*}(\umu;\theta,h)  }
\qquad;\qquad 
\begin{array}{c} 
\displaystyle \alpha=\frac{a_1+1+\theta}{\theta}+M -\frac{N}{\theta} \qquad;\qquad \beta= M-\frac{N}{\theta} \\[.3cm] 
\displaystyle h= \frac{a_1+a_2+1}{2}+\theta M -N
\end{array}
\end{equation}
where the ${\tilde P}^{\star}$ are the so called super BC interpolation polynomials. 
The $\tilde{P}^{\star}$ are the aforementioned polynomials whose values at integer partitions 
compute the non trivial part of the binomial coefficient. The other objects in \eqref{Spaul} are
\begin{equation}
C^0_{\underline{\kappa}}(w;\theta)= \prod_{(ij)\in\underline{\kappa}}(j-1-\theta(i-1 )+w)\qquad;\qquad C^0_{\underline{\kappa}}(a,b;\theta)=C^0_{\underline{\kappa}}(a;\theta)C^0_{\underline{\kappa}}(b;\theta)
\end{equation}
and then $\Pi(\theta=1)=1$. See also \cite[(7.10)]{Aprile:2021pwd}.\\  

The polynomial ${\tilde P}^{*}$ is specified by a number of variables, a Young diagram, 
and an extra parameter that we are calling $h$. Remarkably they are uniquely defined by a 
list of properties, and in particular a vanishing property when evaluated on 
partitions: $\tilde{P}^{\star}_{\ulmb}(\umu;\theta,\gamma)=0$ if $\umu$ does not contain $\ulmb$.\\

The binomial coefficient is stable, which means that it does not depend on the number 
of variables that we input to specify the $\tilde{P}^{*}$. In other words, this input is just an 
intermediate step of the computation: the final result does not depend on it! Thanks to this 
independence, we can simplify our task here and consider ordinary $\tilde{P}^{(*)}$ of $d$ 
variables, that for $\theta=1$, admit a very simple determinantal formula \cite[(7.3)]{Okounkov_Olshanski}:
\begin{equation}\label{PstarOkounkov}
{\tilde P}^{\star}_{\ulmb}(x;\theta=1,h) = \frac{ \ \ \ \ \ \ \det \big( {\cal M}_{ij}\big)_{1\leq i,j\leq d}}{\displaystyle \!\!\!\!\prod_{1\leq i<j\leq d}\!\!\!\!\!\left((x_i-i+h)^2 -(x_j-j+h)^2\right)}\qquad;\qquad
{\cal M}_{ij}= \prod_{k=1}^{\lambda_j-j+d} \left( (x_i-i+h)^2 -(k-1-d+h)^2 \right)
\end{equation}
The value of $d$ can be taken to be the number of rows of $\umu$ when we 
evaluate ${\tilde P}^{\star}_{\ulmb}(\umu)$. The normalization is known in closed form,
\begin{equation}
\tilde{P}^{\star}_{\umu}(\umu;\theta,\gamma)=\prod_{(i,j)\in\umu}\big(1+\mu_i-j+\theta(\mu'_j-i)\big)\big(2\gamma-1+\mu_i+j-\theta(\mu'_j+i)\big)
\end{equation}
One can check from the determinant that $\tilde{P}^{\star}_{\ulmb}(\umu;\theta,\gamma)=0$ if $\umu$ 
does not contain $\ulmb$. Thus the $(S_{\gamma})^{\umu}_{\ulmb}$ are determined, 
and hence the super Jacobi polynomials through \eqref{jacobiinjack}.
We refer to \cite[appendix C]{Aprile:2021pwd} for more details, and a concise 
introduction about the case $\theta\neq 1$, which is relevant for three, five, and six-dimensional (S)CFTs.\\

%================================================================================
\subsection{$M$aster coefficients for all free theories}
\label{sec:all free}

We understood that the $M_{k,\gamma,\ulmb}$ for all free theories are computed by
\begin{equation}
M_{k,\gamma,\ulmb}=\ J_{\ulmb}^{(\beta,0)}(1^k,0^{\beta-k};p_{-},p_{+})  \qquad 0\leq k\leq \beta 
\end{equation}
and by 
\begin{equation}\label{SJACOBI2}
M_{k,\gamma,\ulmb}=\left\{\begin{array}{cc} 
J_{\ulmb}^{(\beta-k,-k)}(0^{\beta-k},1^{-k}) & k<0 \\[.3cm]
%J_{\ulmb}^{(\beta,0)}(1^k0^{\beta-k})  &0\leq k\leq \beta \\[.3cm]
J_{\ulmb}^{(k,k-\beta)}(1^k,0^{k-\beta}) & k>\beta
\end{array}\right.
\end{equation}
where $\beta=\frac{1}{2}(\gamma-\gamma_{\min})$. Here we will give a determinant 
formula for the latter by using our results for the super Jacobi polynomials in \eqref{SJACOBI}. 
One point to make about \eqref{SJACOBI} is that the formula  is naively singular at coincident 
values of the $X,Y$ variables, but of course we are just evaluating a polynomial and so the 
result must be finite. The way to proceed is to absorb the singular pieces of the Vandermonde 
into the numerator, and manipulate rows and column of the determinant in order to take a manifestly finite limit.\\

Let us see how the singular Vandermonde is resolved in an example. Say there are $k$ 
variables $X$ or $Y$ to be evaluated at the same value. Let us take $Y$ for concreteness. 
This means that there are $\frac{1}{2}k(k-1)$ factors $Y_a-Y_b$ in the Vandermonde that are 
singular. Organize these into sequences as follows. The first sequence is given by the  $k-1$ 
pairs $\prod_{i=1}^{k-1}(Y_i-Y_k)$. The second sequence is given by the $k-2$ pairs 
$\prod_{i=1}^{k-2}(Y_i-Y_{k-1}) $. Etc\ldots. The last sequence is made by the one pair $(Y_1-Y_{2})$. 
A determinant does not change if we add/subtract rows, therefore consider the following moves. 
Starting from the first sequence replace the row $r_i$ of the determinant with $r_i\rightarrow (r_i-r_k)/(Y_i-Y_k)$. 
This removes all the singular factors involving $Y_k$. Change sequence, and repeat the 
procedure (starting from the updated rows) until the last sequence is gone. At this point, the 
simplest way to take the limit of coincident values is to Taylor expand around $Y_k$. As a result, 
row-one is proportional to the $(k-1)$-th derivative of the original $r_1$, then row-two is 
proportional to the $(k-2)$-th derivative of the original $r_2$, etc\ldots. The last row is unchanged 
and in particular has no derivatives. 
In this example we used rows, but the very same procedure can be applied to columns as well. 
In fact, the combination of both is what we have to use when dealing with the evaluation of the super 
Jacobi polynomials in \eqref{SJACOBI2}, because the Young diagram in this case is always atypical, 
and therefore the determinant does not factorize. In particular, the matrix $R_{ij}$ is always active for 
atypical diagrams, but crucially, it is non singular since $X\neq Y$.\\

Since the coincident values are either $0$s or $1$s, we will need formulae 
for the derivatives of the one-variable Jacobi polynomials. These are
\begin{align}\label{derJacobi_onevar}
\frac{d^q}{dy^q}J_{[\lambda]}(1,a_1,a_2)= \frac{\lambda!(a_1+a_2+q+\lambda)!(a_2+\lambda)!}{(\lambda-q)!(a_1+a_2+2\lambda)!(a_2+q)!}\qquad;\qquad
\frac{d^q}{dy^q}J_{[\lambda]}(0,a_1,a_2)= (-)^{q+\lambda}
\frac{d^q}{dy^q}J_{[\lambda]}(1,a_2,a_1)
\end{align}
The two derivatives are related to each other, but note the reverse order of $a_{i=1,2}$ when evaluating in zero.\\ 

The determinant formula that gives the evaluation formula of the super Jacobi polynomials in \eqref{SJACOBI2}, is,
\begin{equation}\label{SJACOBI3}
J^{(\mathsf{M}|\mathsf{N})}_\ulmb(X,Y)\Bigg|_{\eqref{SJACOBI2}}
=\  
%(\X-\Y)^{(MN)} 
\frac{(-)^\Sigma}{\prod_{i=1}^{\mathsf{M}-1} i! \prod_{j=1}^{\mathsf{N}-1} j!} \!\det\!\left( \begin{array}{cc} 
R_{ij}& \nabla E_{\ulmb}(Y) 
\\[.2cm]
\nabla S_{\ulmb}(X) & 0\ \ \ \end{array} \right)  \\[.2cm]
\end{equation}
where the values of $\mathsf{M},\mathsf{N},X,Y$ should be taken accordingly to \eqref{SJACOBI2}, and where
\begin{equation}
(-1)^\Sigma=(Y-X)^{\mathsf{M}\mathsf{N}}(-1)^\sigma\qquad;\qquad
R_{ij}=(\mathsf{M}-i+\mathsf{N}-j)!\frac{(-)^{\mathsf{M}-i} }{(Y-X)^{\mathsf{M}-i+\mathsf{N}-j+1}}
\end{equation}
with 
\begin{align} 
\big[ \nabla E_{\ulmb}(0;p^-,p^+)\big]_{il}&=(-1)^{\mathsf{M}-i+\hat{\lambda}_l}
\big[ \nabla E_{\ulmb}(1;p^+,p^-)\big]_{il} \\[.2cm]
\big[ \nabla E_{\ulmb}(1;p^-,p^+)\big]_{il}&= \frac{
(\hat{\lambda}_l)!(\hat{\lambda}_l+\mathsf{M}-i+\hat{p})!(\hat{\lambda}_l+p_+)!}{(\hat{\lambda}_l-\mathsf{M}+i)!(2\hat{\lambda}_l+\hat{p})!(\mathsf{M}-i+p_+)! } \\[.2cm]
\big[ \nabla S_{\ulmb}(0;p^-,p^+)\big]_{lj}&=(-1)^{\mathsf{N}-j+\hat{\lambda}'_l}\big[ \nabla S_{\ulmb}(1;p^+,p^-)\big]_{lj} \\[.2cm]
\big[ \nabla S_{\ulmb}(1;p^-,p^+)\big]_{lj}&= \frac{
(\hat{\lambda}'_l)!(\hat{\lambda}'_l+\mathsf{N}-j-\hat{p})!(\hat{\lambda}'_l-p_+)!}{(\hat{\lambda}'_l-\mathsf{N}+j)!(2\hat{\lambda}'_l-\hat{p})!(\mathsf{N}-j-p_+)! } %\\[.2cm]
\end{align}
and $\hat{\lambda}'_l=\lambda'_l-\beta-l$, $\hat{\lambda}_l=\lambda_l+\beta-l$. \\

At this point we observe that the manipulations that allow us to write \eqref{SJACOBI3} 
are valid also when $\mathsf{N}=0$, i.e.~when the super Jacobi polynomial reduces to a 
Jacobi polynomial. With some additional work,
this gives a direct proof of the evaluation formulae \eqref{evaluationknown1}-\eqref{evaluationknown2}, 
and  \eqref{Acoeffs}. For the interested reader, the only subtle point is to understand that a matrix whose rows are built from 
$\big[ \nabla E_{\ulmb}(0;p^-,p^+)\big]_{il}$ and
$\big[ \nabla E_{\ulmb}(1;p^-,p^+)\big]_{il}$ falls into the class of VdM-type determinants 
of the Pochhammer type, which can be evaluated as in  \cite[Proposition $1$]{adv_det_calc}.\\

There are, however, two important comments to make when comparing the evaluation 
of super Jacobi in \eqref{SJACOBI3} and our master formula for the scalar case:
\begin{itemize}
\item[1)]
While the master formula \eqref{Acoeffs} takes the form of a sum over quantities that 
are manifestly analytic in spin (multiplied by $(-1)^l$ factors), in the case of super Jacobi 
polynomials the Laplace expansion of the determinant is complicated by the presence 
of the matrix $R_{ij}$, and therefore we will not further manipulate \eqref{SJACOBI3}.
\item[2)] In the master formula \eqref{Acoeffs} the Young diagram has an upper bound 
on the number of allowed rows, which depends on $\beta$, namely $\ulmb=[\lambda_1,\ldots,\lambda_{\beta}]$, 
with $\lambda_i$ that might be vanishing. In the \eqref{SJACOBI3} this is not the case anymore. 
For example, for $\gamma=2$ and $\beta=1$ we can have a Young diagram with two rows.
\end{itemize}

\subsection{Worked example}

We conclude this section with an example. Consider the four point 
function \cite[(6.27)]{Dolan:2000ut} of a dimension three operator made from a free fermion field, 
\begin{align}
\langle O_3 O_3 O_3 O_3\rangle = 
g_{12}^3 g_{34}^3 \Bigg( &\,\frac{1}{4} \Bigg[ -U -\frac{U}{V} + UV+ \frac{U}{V^2}\Bigg] 
-\frac{1}{4} \Bigg[ U^2 +\frac{U^2}{V^2}\Bigg] + \Bigg[ U^3 -\frac{1}{4}\frac{U^3}{V}-\frac{1}{4} \frac{U^3}{V^2} +\frac{U^3}{V^3}\Bigg] +\frac{1}{4}\frac{U^4}{V^2} \Bigg)\,.
\end{align}
The terms $UV+U/V^2$ will require the evaluation of super Jacobi polynomials. 
In any case the conversion from Young diagrams to dimension and spin is
\begin{equation}
\Delta = \lambda_1+\lambda_2+\gamma \,, \qquad l =\lambda_1-\lambda_2 \,.
\end{equation}
Below we compare our block decomposition with that in \cite[(6.31)-(6.32)]{Dolan:2000ut}. 
For the first two twists we find
\begin{center}
\def\arraystretch{1.5}
\centering
\begin{tabular}{|c|c|c|}
%\hline 
% 
\hline\hline
$\tau=2$  &  $\rule{0pt}{.6cm} \frac{l!(l+1)!}{4(2l-1)!}$  &  $\frac{1}{4}\Big[ -M_{0,2,[l]} -M_{1,2,[l]}+M_{-1,2,[l]}+M_{2,2,[l]}\Big]$  \\[2ex] 
\hline
$\tau=4$  & \rule{0pt}{.6cm} 0  &  $\frac{1}{4}\Big[M_{-1,2,[l+1,1]}+M_{2,2,[l+1,1]}\Big]$  $-\frac{1}{4}\Big[ M_{0,4,[l]} +M_{2,4,[l]}\Big]$ \\[2ex] 
\hline
\end{tabular}
\end{center}
Note the presence of $M_{-1,2}$ and $M_{2,2}$ in both rows, which are crucial to 
obtain the correct result. The zero result at $\tau=4$ generalizes to higher twists. In fact, the following identity holds
\begin{equation}
\Big[M_{-1,2,[l+t-1,t-1]}+M_{2,2,[l+t-1,t-1]}- M_{0,4,[l+t-2,t-2]} -M_{2,4,[l+t-2,t-2]}\Big]=0\qquad t\ge 2
\end{equation}
Therefore we will not sum these contributions to higher twists.
Then for $\tau=6$ we find
\begin{equation}
\frac{(l+2)!(l+3)!(l^2+5l +3)}{4(2l+3)!}= 
\Big[ M_{0,6,[l]} -\frac{1}{4}M_{1,6,[l]}-\frac{1}{4}M_{2,6,[l]}+M_{3,6,[l]}\Big]
\end{equation}
and the result for $\tau\ge 8$ is 
\begin{equation}
 \Big[ M_{0,6,[l+\frac{\tau}{2}-3,\frac{\tau}{2}-3]} -\frac{1}{4}M_{1,6,[l+\frac{\tau}{2}-3,\frac{\tau}{2}-3]}-\frac{1}{4}M_{2,6,[l+\frac{\tau}{2}-3,\frac{\tau}{2}-3]}+M_{3,6,[l+\frac{\tau}{2}-3,\frac{\tau}{2}-3]}\Big]+ \frac{1}{4}M_{2,8,[l+\frac{\tau}{2}-4,\frac{\tau}{2}-4]} 
\end{equation}
which equals to $\frac{(l+\frac{\tau}{2}-1)!(l+\frac{\tau}{2})!(\frac{\tau}{2}-1)!(\frac{\tau}{2}-2)!}{8(\tau-5)!(\tau+2l-3)!}((l+1)(\tau+l-2)+(-1)^{\frac{\tau}{2}})$. 
We thus find perfect agreement with the known results.\\ 

We also checked the case of the vector field in \cite[(6.34)]{Dolan:2000ut} 
and found perfect agreement.

%%%%%%%%%%%%%%%%%%%
\bibliographystyle{apsrev4-2}

\begin{thebibliography}{99}
\bibitem{footnote1}
{The precise 
form of $\cO_{\gamma,\ulmb}$ is obtained by taking all possible $b_{12}$ Wick contractions. 
Similarly for $\cO'_{\gamma,\ulmb}$ and $b_{34}$.}
\bibitem{footnote2}
{There are also $\frac{1}{4}$-BPS operators 
which naturally occur in families with the semishort operators 
and therefore we won't distinguish them}
\end{thebibliography}

\end{document}